\documentclass[showkeys,%
preprint%
amsmath,amssymb,%
aps,%
pre,%
groupedaddress,%
superscriptaddress,%
a4paper,%
floatfix,%
eprint,%
preprintnumbers,%
eqsecnum,%
nofootinbib,%
]{revtex4-2}

\usepackage[utf8]{inputenc}
\usepackage{fontenc}
\usepackage{amsmath}
\usepackage{subfigure}
\usepackage{color,xcolor}
\usepackage{multirow}
\usepackage{orcidlink}
\usepackage{physics}
\usepackage{dcolumn}
\usepackage{bm}
\usepackage{graphics,graphicx} 
\usepackage{natbib}
\usepackage{xcolor,color}
\usepackage[]{hyperref}
\hypersetup{colorlinks=true,
	breaklinks,
	hyperfootnotes,
	linkcolor={red},
	final,
	urlcolor={purple},
	citecolor=cyan,
	breaklinks=true}

\usepackage{booktabs}
\usepackage{verbatim}

\graphicspath{{Figures/}} 
\begin{document}
\normalsize
\title{Formation and Localization of Four-wing Attractor in Phase space}

\author{Tanmayee Patra\,\orcidlink{0009-0007-5555-0068}}
\email{tanmayeepatra1995@gmail.com}
\affiliation{%
Department of Physics and Astronomy, National Institute of Technology Rourkela, Odisha-769008, India}%

\author{Biplab Ganguli\,\orcidlink{0000-0003-2583-5752}}%
\email{biplabg@nitrkl.ac.in}
\affiliation{%
Department of Physics and Astronomy, National Institute of Technology Rourkela, Odisha-769008, India}%
.
\begin{abstract}
	
A chaotic attractor is formed in a finite region in phase space by the long-term trajectory of a three or higher-dimensional dissipative system. The attractor is a fractional-dimensional geometry, whose dimension is a fraction but less than the dimension of the phase space. The geometry of an attractor can be as complex as a multi-wing geometry. The emergence and confinement of such a complex geometrical attractor can be understood by the Nambu mechanics without numerically solving the governing equations of the dynamics. In this article, we show that the four-wing geometry of an attractor appears in the phase space by the intersection of two energy-like Hamiltonian functions. We further show that the dynamical equations require the localization range of these surfaces so that their intersection is confined to a certain region of the phase space. We analytically find the required conditions based on the system parameters for the localization of the attractor.

\end{abstract}

\maketitle
\pagestyle{plain}
\section{\label{sec1}Introduction}
The most basic technique of studying any chaotic attractor is to study the phase space portrait \cite{RevModPhys.53.655} and the geometrical shape/structure of such an attractor gives a clear indication of the nature of the dynamics. Phase-space trajectories are indispensable for analyzing chaotic attractors. Their significance can be highlighted in three key aspects: (i) they unveil the intricate `shape' of chaos, offering a geometric lens through which we can visualize the system's evolution, (ii) they enable the precise quantification of chaotic measures, such as Lyapunov exponents and fractal dimensions, which are essential for deeper insights into chaotic dynamics, (iii) they serve as a powerful tool to distinguish chaos from mere noise or periodic behaviour. 

The term `wings' when applied to various attractors, it encapsulates the unique and striking shapes of their chaotic attractors in phase space.  The wings that characterize chaotic attractors are not merely decorative; they reveal intricate structures akin to lobes or butterfly-like shapes, vividly depicted in their phase portraits. The topology of the attractor showcases an impressive diversity, with configurations that can range from just one wings to a multitude \cite{2014-8-080505,doi:10.1142/S0218127418500451, ZHANG2018793, 10.3389/fphy.2022.927991, SAHOO2022112598, fractalfract8070417}. As articulated by Guoyuan Qi \textit{et al.} in their research \cite{wang20093, qi2008four}, the concept of a wing encompasses a profound understanding of multi-wing attractors. These chaotic systems manifest complex dynamical behaviours around multiple equilibrium points. The phase portraits of multi-wing attractors resemble an array (multiple) of wings or lobes, arranged both symmetrically and asymmetrically around a central point \cite{10324344, SAHOO2022112598}. The creation of these wings arises from the nonlinear folding and stretching of trajectories, which invite exploration and inquiry. While these systems often occur around one or a few equilibrium points, their swirling trajectories produce multiple distinct wing-like structures, pushing the boundaries of our understanding and encouraging further investigation into their complex dynamics.

 The common approach to studying chaotic attractors in phase space is to directly solve the governing dynamical equations. The convergence time of trajectories to the attractor depends on the choice of initial conditions, parameter values and the dynamics. The numerical approach is non-trivial in the case of chaotic dynamics, for which two crucial points should be kept in mind. Firstly, one must have a good idea of the basin of attraction. Secondly, the system has to be run for a long time and this is because of the aperiodic nature of chaotic dynamics.  Along with this, there are a few more rigorous numerical techniques, such as bifurcation diagrams \cite{strogatz2007nonlinear}, Lyapunov exponents \cite{doi:10.1142/S021797929100064X}, Poincaré maps \cite{shahhosseini2023poincare}, fractal dimensions \cite{bakri2025note}, power spectra \cite{PhysRevLett.58.1699}, topological templates \cite{rosalie2014toward}, linking numbers \cite{PhysRevE.49.4693} etc., that are implemented to study the nature of dynamics, which lead us to get various affirmative conclusions regarding the characterization of chaotic attractors. An important research by N. Tuffillaro \textit{et al.} \cite{tufillaro1992experimental} and R. Gilmore \cite{gilmore1998topological}, serve as an essential resource for delving into the complexities of chaos and they offer a compelling exploration of the mathematical principles underlying non-linear dynamics, complemented by a topological approach that enhances the analysis of chaotic behaviour.
 
  In other words, we can state that quantitative studies of chaotic dynamics rely on the numerical solutions of governing equations, enabling the creation of insightful phase-space portraits. The time series obtained from these solutions serves as the foundation for various characterization tools, allowing us to explore the complexities of chaotic behaviour. By calculating the Lyapunov exponent, we can identify regions of periodic and chaotic dynamics, while bifurcation diagrams generated through Poincaré maps vividly illustrate pathways to chaos.
  Apart from extensive numerical study of chaotic dynamics and attractor based on tools mentioned above, there are studies, focussing on the geometrical and topological properties of attractors also. Extensive research into the topological properties \cite{PhysRevLett.91.134104, rosalie2014toward, rosalie2015systematic} of attractors has led to the development of tools that quantify these properties using numerical solutions. Creating attractor templates and calculating linking numbers help us understand the phenomena of folding and stretching in fractional-dimensional spaces. At the core of these methodologies, the identification of unstable periodic orbits (UPOs) plays an important role, which are derived from the first return map of the time series. This allows for the construction of detailed templates that provide clear topological representations. These innovative tools have been applied to a range of attractors, from the Rössler attractor \cite{ rosalie2016templates} to more complex systems like the Lorenz attractor \cite{rosalie2025structure}. 
  J. Guckenheimer and R. Williams \cite{guckenheimer1979structural, williams1979structure} utilized rigorous analytical tools involving branched manifolds to examine the geometric structure of the Lorenz attractor.   Shilnikov \cite{kazakov2021leonidshilnikovmathematicaltheory} developed highly sophisticated mathematical tools to study stability of the boundary of an attractor which predicts first return map and bifurcation. Using his mathematical theory, he developed a new class of chaos known as `Shilnikov chaos'. The concept of Shilnikov chaos was explored by T. Zhou \textit{et al.} \cite{zhou2005vsi} in the generalized Lorenz canonical form of the dynamical systems. In 1992, V. V. Bykov and A. L. Shilnikov \cite{bykov1989boundaries} studied the boundaries of existence of the Lorenz chaotic attractor. Later, M. Capinski, D. Turaev, and P. Zgliczynski \cite{capinski2018computer} proved the Lorenz attractor's existence in the Shimizu-Morioka system using Shilnikov's criterion.

 To understand the localization of Lorenz attractor and the geometry of the boundary of the attractor,  Doering and Gibbon \cite{doi:10.1080/02681119508806207} consider various types of surfaces and provide localization of the boundary of Lorenz attractor within such surfaces. The idea behind this is to find time evolution of these surfaces using the dynamical equations of the Lorenz equations and show that they are either attracting or repelling surfaces. Though they could provide that these surfaces attract/repel trajectories, but it does not prove the geometry of the boundary of Lorenz attractor.
 
 An alternative, yet under-explored approach for exploring the geometry of the boundary of chaotic attractors and their localization, lies in the realm of Nambu mechanics \cite{Takhtajan_1994}. Nambu mechanics is a generalization of classical Hamiltonian mechanics proposed by Yoichiro Nambu in 1973 \cite{PhysRevD.7.2405}. It stands at the forefront of mathematical physics research, offering profound insights into systems, characterized by multiple conserved quantities and intricate algebraic structures \cite{guha2002applications, 10.1093/ptep/ptab075}. An illustration of its application is found in the Euler equations for a rigid body. These equations can be  reformulated through Nambu mechanics, utilizing two Hamiltonians—energy and the square of angular momentum. This powerful representation can be concisely expressed as \(\frac{dL}{dt} = \{L, H, C\}_{NB}\), with \(C = \frac{1}{2}L^2\) \cite{Axenides_2010}. The theory of Nambu mechanics is well suited for conservative systems which may have several conservative quantities. Because of its applicability for conservative systems, it did not attract attention of the nonlinear dynamics community until P. Nevir and R. Blender \cite{nevir1994hamiltonian} first applied Nambu mechanics to the Lorenz attractor, setting the stage for further advancements by M. Axenides and E. Floratos \cite{Axenides_2010}, Z. Roupas \cite{Roupas_2012}, and W. Mathis and R. Mathis \cite{mathis2014dissipative}. They show that, if the vector flow of a dynamical system is splitted into two separate parts, a non-dissipative and a dissipative, Nambu mechanics can be applied to the non-dissipative part to describe the geometry of an attractor using several  energy-like Hamiltonian functions. This method demonstrates that the geometrical shape of an attractor can be understood solely from the functional forms  of vector fields, eliminating the need for numerical time series data. As a result, it can be extended to explain the formation of intricate multi-lobe structures. This approach not only enhances our understanding of the system behaviours in phase space but also exemplifies an innovative geometric perspective for dissecting any chaotic attractor. 
  
 Within this framework of the mechanics, it is possible to define energy-like functions, which completely describe the dynamics as well as the nature of the boundary of an attractor. Roupas  \cite{Roupas_2012} showed that surfaces used by Doering and Gibbon \cite{doi:10.1080/02681119508806207} can be obtained by linear combinations of these Hamiltonian functions and the localization of an attractor can be proved by these surfaces.

 Aside from the application of Nambu mechanics to Lorenz and Rössler attractors \cite{Axenides_2010} which have geometry of two and one wing respectively, research into additional attractors that can substantiate the significance of this geometric perspective within Nambu mechanics remains limited. In this paper, we enhance Nambu mechanics by applying it to a more complex 3-dimensional autonomous chaotic attractor known as the four-wing attractor, which was first introduced by J. Lu \textit{et al.} in 2004 \cite{lu2004new} and further explored by G. Qi \textit{et al.} in 2008 \cite{qi2008four} and Z. Wang \textit{et al.} in 2009 \cite{wang20093}. The properties of a four-wing system with symmmetry group $V_4$ was explored by C. Letellier and R. Gilmore in \cite{letellier2007symmetry}. The motivation of our research is only to demonstrate the applicability of Nambu mechanics as a general theoretical framework which can address the issue of localization of an attractor in phase space and answer the cause of formation of complex geometry of an attractor. We therefore intentionally selected a well studied known system \cite{wang20093} that exhibits four-wing(multi-lobe) geometry in phase space for a fixed value of various system parameters. We demonstrate that Nambu mechanics can generate more complex Hamiltonian like surfaces which could also generate higher lobes attractor  than only Lorenz like two lobes attractor. Therefore, there are two important questions, that may be answered : Firstly, how the different wing like structures formed in the phase space, which can be found from the vector field alone without solving the underline dynamical equations? Secondly, how the energy like functions are responsible for localization of the attractor? We further show that these energy like surfaces are not only attracting or repelling, but we demonstrate first time that intersection of pair of surfaces alone determines the localization and geometry of the attractor. While applying Nambu Mechanics, Roupas followed the similar justification as Doering and Gibbon \cite{doi:10.1080/02681119508806207} and argued that Nambu surfaces can be attracting surfaces to achieve localization. In contrast to this, we demonstrate that the localized geometry(of attractor) is formed due to intersection than attracting surfaces alone. We further find the condition on system parameter for localization using extremum condition on the surfaces.
 
The four-wing attractor is a type of strange attractor that appears in certain chaotic systems.  Although it may not be as extensively studied as the Lorenz or Rössler attractors, various research articles have explored its properties and significance \cite{khan2020behavioral, hu2009hyperchaotic}. These articles provide valuable insights into the formation, characteristics, and relevance of four-wing attractors in chaotic systems. 

In section \ref{sec2}, we provide a brief introduction to the concept of Nambu mechanics. In section \ref{sec3}, we present a complex four-wing chaotic attractor studied by Z. Wang \textit{et al.} \cite{wang20093}, which can be generated from a set of three first-order autonomous ordinary differential equations. In section \ref{sec4}, we discuss the concept of Nambu surfaces in relation to the four-wing strange attractor and explain how this attractor emerges from the intersection of two Nambu surfaces. More precisely, subsection \ref{sec4a} demonstrates the construction of two Nambu functions derived from the non-dissipative part of the system. Additionally, in subsection \ref{sec4b}, we explore how different Nambu functions can be transformed from the parent Nambu doublet to describe identical dynamics. In subsection \ref{sec4c}, we extend our analysis to include dissipation in the phase space, showing that dissipative Nambu dynamics can geometrically interpret the complete chaotic orbits of the four-wing attractor. Moreover, we discuss various conditions for identifying the localization boundaries of the attractors in subsection \ref{sec4d}.  Lastly,  we summarize our results in section \ref{sec5}.

\section{\label{sec2}Nambu Mechanics}

Nambu mechanics generalizes the conventional Poisson bracket formalism by integrating multiple conserved quantities, paving the way for complex higher-dimensional phase-space dynamics. The key features that make Nambu mechanics a groundbreaking approach are: (i) Nambu-Poisson bracket representation : Instead of the usual conventional Poisson bracket $\pb{U}{V}_{PB}$ in Hamiltonian mechanics, Nambu mechanics introduces a triple bracket (or more generally, n-ary bracket) defined as $\qty{U, V, W}_{NB} = \vec{\nabla} U \cdot (\vec{\nabla}V \times \vec{\nabla} W)$ where, $U$, $V$, and $W$ are functions on phase space, and $\vec{\nabla}$ represents the gradient in three-dimensional space $(x, y, z)$. This formulation enhances the dynamical framework of physical systems, (ii) Multiple Hamiltonians : Instead of relying on a single Hamiltonian $H$, Nambu mechanics harnesses two or more Hamiltonians ($H_1$, $H_2$ or more), to drive time evolution. This flexibility opens up new avenues for understanding dynamics in such a way that traditional methods cannot, (iii) Volume-preserving flow : Liouville theorem is the guiding principle for Nambu mechanics. This principle ensures that the dynamics of the system preserve volume in phase space. Just as Hamiltonian mechanics preserves area in phase space, Nambu mechanics preserves phase space volume. In summary, Nambu mechanics isn't just a theoretical enhancement; it is a paradigm shift that offers a deeper understanding of a higher dimensional complex dynamical systems.

 In Hamiltonian mechanics, the Poisson bracket describes the dynamics involving a pair of canonical variables. Nambu mechanics extends this concept by involving Poisson-like brackets with three or more canonical variables \cite{PhysRevD.7.2405}. This generalization is particularly useful for describing dynamics in phase spaces of three dimensions or more. In Hamiltonian mechanics, the evolution of any canonical variable is determined by
\begin{equation}
\dot{x_{i}}=\pb{x_{i}}{H}_{PB}
\end{equation}
and it requires one constant of motion, $H$.

In Nambu mechanics, the dynamical equation of n canonical variables requires n-1  Hamiltonian like functions $H_{1},H_{2},...H_{n-1}$ such that 
\begin{equation}
\dot{x_{i}}=\qty\big{x_{i},{H_{1}},{H_{2}},...{H_{n-1}}}_{NB} =\epsilon_{ijk...l}\partial_{j} H_{1}\partial_{k} H_{2}...\partial_{l} H_{n-1}=\sum_{jk...l}\epsilon_{ijk...l}\frac{\partial H_{1}}{\partial x_{j}} \frac{\partial H_{2}}{\partial x_{k}}...\frac{\partial H_{n-1}}{\partial x_{l}}
\end{equation}
     
For  3-canonical variables, we get     
\begin{equation}\label{nam}
\dot{x_{i}}=\qty\big{x_{i},{H_{1}},{H_{2}}}_{NB} =\epsilon_{ijk}\partial_{j} H_{1}\partial_{k} H_{2}=\sum_{jk}\epsilon_{ijk}\frac{\partial H_{1}}{\partial x_{j}} \frac{\partial H_{2}}{\partial x_{k}}=(\vec\grad{H_{1}}\cross\vec\grad{H_{2}})_i
\end{equation}

The Nambu-Poisson bracket for equation (\ref{nam}) is given by

\begin{equation}\label{nam1}
	\qty\big{x_{i},{H_{1}},{H_{2}}}_{NB}=\qty\big{x_{i},H_{1}}_{H_{2}}=\epsilon_{ijk}\partial_{j} H_{1}\partial_{k} H_{2} =(\vec\grad{H_{1}}\cross\vec\grad{H_{2}})_i
\end{equation}

In this Nambu-Poisson notation, the Hamiltonian outside the Nambu bracket denotes a 2D-Euclidean space, and the Hamiltonian inside the Nambu bracket describes the dynamics of the system on that 2D-Euclidean space. In notation $\qty\big{x_{i},H_{1}}_{H_{2}}$, the position and role of $H_{1}$ \& $H_{2}$ can be interchanged.

Therefore, from equation (\ref{nam} and \ref{nam1}) we get three variables dynamical equation, which is given by
\begin{equation}\label{NDeq}
	\dot{\vec{x}} = \vec v=\vec\grad{H_{1}}\cross\vec\grad{H_{2}}
\end{equation}

The equation (\ref{NDeq}) represents the dynamics at the intersection of $H_1$ and $H_2$. Here, the gradients of $\vec\grad{H_{1}}$ and $\vec\grad{H_{2}}$ are normal to the surfaces $H_{1}$ and $H_{2}$ respectively and the cross-product ($\vec\grad{H_{1}}\cross \vec\grad{H_{2}}$) gives a vector perpendicular to both the normals $\vec\grad{H_{1}}$ and $\vec\grad{H_{2}}$. Therefore, the cross product vector should be tangent to both the surfaces $H_{1}$ and $H_{2}$ simultaneously, indicating the curve of intersection of both the surfaces. Hence, the trajectory must lie on both the surfaces simultaneously. 

$H_{1}$ \& $H_{2}$ serve as conserved quantities because  of Liouville theorem and therefore the velocity field is divergence-free, preserving the volume of phase space. This fundamental principle highlights that Nambu mechanics demands for two Hamiltonian surfaces for any three-dimensional system describing the full dynamics. We can succinctly represent this as $h= (H_{1},H_{2})$, illustrating that these two Hamiltonians form a  Nambu doublet. One Hamiltonian represents a two-dimensional phase space in $R^{3}$, while the other governs the time evolution of the system within this two-dimensional phase space. This duality not only enhances our understanding of the system but also underscores the uniqueness of Nambu mechanics \cite{Takhtajan_1994, 2009}. 

Moreover, the equation (\ref{NDeq}) remains invariant if $H_1$ and $H_2$ are transformed to new Hamiltonians $H'_1$ and $H'_2$, subject to the condition 
\begin{equation}
\left|\frac{\partial \left(H'_{1},H'_{2}\right)}{\partial \left(H_{1},H_{2}\right)}\right|=1
\end{equation}

 Here, the left-hand side represents the determinant of the transformation matrix (or Jacobian matrix), while $H'_1$  \&  $H'_2$ are functions of $H_1$ \& $H_2$. This indicates that the transformation is governed by a $2 \cross 2$ matrix in the $H_1$ and $H_2$ space. Consequently, we can transform the Nambu doublet ($H_1, H_2$) by appropriately selecting a transformation matrix, subject to the  condition that the determinant of the transformation matrix is equal to one. This transformation is called a canonical transformation. By applying this transformation law, we can generate a variety of transformed Nambu surfaces from the parent/original Nambu doublet, unlocking new avenues for exploration and understanding in this field. Hence, the transformation law of the Nambu doublet ($H_1, H_2$) is given by
\begin{equation}
	\begin{split}
		(\vec\grad{H'_{1}}\cross\vec\grad{H'_{2}})_i&=\epsilon_{ijk}\partial_{j}{H'_{1}}\partial_{k}{H'_{2}}\\&
		=\epsilon_{ijk}\qty\Bigg(\pdv{H'_{1}}{H_{1}}\partial_{j}{H_{1}}+\pdv{H'_{1}}{H_{2}}\partial_{j}{H_{2}})\qty\Bigg(\pdv{H'_{2}}{H_{1}}\partial_{k}{H_{1}}+\pdv{H'_{2}}{H_{2}}\partial_{k}{H_{2}})\\&
		=\epsilon_{ijk}\qty\Bigg(\pdv{H'_{1}}{H_{1}}\pdv{H'_{2}}{H_{2}}-\pdv{H'_{1}}{H_{2}}\pdv{H'_{2}}{H_{1}})\partial_{j}{H_{1}}\partial_{k}{H_{2}}\\&
		=\left|\frac{\partial \left(H'_{1},H'_{2}\right)}{\partial \left(H_{1},H_{2}\right)}\right|\quad \left(\vec\grad{H_{1}}\cross\vec\grad{H_{2}}\right)_i  
	\end{split}
\end{equation}

Since the dynamics is unique, all possible transformed doublets represent the same trajectories in phase space. As a result, we expect the intersection of surfaces corresponding to any doublet (pair of Hamiltonians)  to remain consistent. We will demonstrate this property in section \ref{sec4b}.

\subsection{\label{sec2A}Application of Nambu mechanics to chaotic system}
The classic example of application of Nambu mechanics is the dynamics of a free rigid body \cite{Axenides_2010, PhysRevD.7.2405} which can be powerfully captured through the Nambu bracket, and incorporates two essential conserved quantities - energy and the square of angular momentum. But free rigid body is a conservative system, means it has no dissipation part. On the other hand  chaotic systems, like the Lorenz and Rössler attractors are dissipative. We know that the Nambu theory is a framework explicitly tailored for systems, preserving phase-space volume i.e., systems with $\vec \grad \cdot {\vec{v}}=0$. Therefore, dissipative chaotic systems defy the Nambu mechanics framework. In spite of this restriction, it is possible to apply Nambu theory to the Lorenz \cite{Roupas_2012} and Rössler \cite{Axenides_2010} dynamical systems. This is possible because the velocity vector field in these systems can be splitted into a sum of two integral components: a conservative non-dissipative part ($\vec v_{ND}$) and a non-conservative dissipative part ($\vec v_{D}$). In that case, Nambu theory is applied to the conservative part only and thereby giving a completely new interpretation to the dynamics. It turns out that the conservative part($\vec v_{ND}$) of these systems  have multiple conserved quantities, resonating perfectly with Nambu's multi-Hamiltonian formalism. Therefore, this approach not only solidifies our understanding of chaotic system dynamics through the Nambu bracket but also reveals the intricate structure underlying chaotic behaviour. Meanwhile, the $\vec v_{D}$ component is essential for generating the rich chaotic dynamics, driven by dissipation, which significantly impacts the evolution of the resulting Nambu Hamiltonians. 
\subsection{\label{sec2B}Chaotic vector flow decomposition methodology }

As required by the Nambu mechanics, $\vec v_{ND}$ part of the decomposition must be divergenceless, i.e., $\vec \grad \cdot{\vec v_{ND}}=0$. It turns out that for the  Lorenz \cite{Roupas_2012} and Rössler \cite{Axenides_2010} dynamical systems, the other part(dissipative, $\vec v_{D}$) is irrotational, means satisfy the condition of $\vec \grad \cross \vec v_{D}=0$. In other words, we can say that the decomposition of the velocity vector field must satisfy the Helmholtz-Hodge decomposition \cite{6365629}. This property of a velocity field not only simplifies our understanding but also enhances our ability to analyze complex dynamics with precision. Therefore, Helmholtz-Hodge decomposition can be expressed as :
\begin{equation}
	\vec{v}=\vec \grad \cross\vec{A}+\vec\grad{D} \quad\Rightarrow \vec{v}=\vec v_{ND}+\vec v_{D}
\end{equation}
Hence,
\begin{equation}
	\label{eq3}
	\begin{cases}
		\vec v_{ND}=\vec\grad\cross{\vec{A}}\quad\Rightarrow\vec\grad\cdot{\vec v_{ND}}=\vec\grad\cdot{(\vec\grad\cross{\vec{A}})}=0 \quad \text{($\because$ Divergence of curl of a vector = 0)}\\
		\vec v_{D}=\vec\grad{D}\quad\Rightarrow \vec\grad\cross\vec v_{D}=\vec\grad\cross\vec\grad{(D)}=0 \quad \text{($\because$ Curl of gradient of a scalar = 0)}
	\end{cases}\,
\end{equation}
where,  $\vec{A}$ and D represents the vector field and  scalar field respectively.\\

Further, equation (\ref{eq3}) can be rewritten as :
\begin{equation}
	\label{eq4}
	\begin{cases}
		\vec v_{ND}=\vec\grad\cross{\vec{A}}= \vec\grad{H_{1}}\cross \vec\grad{H_{2}}\quad\Rightarrow\vec\grad\cdot{\vec v_{ND}}=\vec\grad\cdot{(\vec\grad{H_{1}}\cross \vec\grad{H_{2}})}=0 \\
		\vec v_{D}=\vec\grad{D}\quad\Rightarrow \vec\grad\cross\vec v_{D}=\vec\grad\cross\grad{(D)}=0 
	\end{cases}\,
\end{equation}
where, \quad $\vec\grad\cross{\vec{A}}$ = $\vec\grad{H_{1}} \cross \vec\grad{H_{2}}\quad\Rightarrow \vec{A}= H_{1} \, \vec\grad{H_{2}}$ (see Appendix \ref{AA} for derivation). Finally we can rewrite the flow vector of the system as 
\begin{equation}
\vec{v}=\vec v_{ND}+\vec v_{D}=(\vec\grad{H_{1}}\cross \vec\grad{H_{2}})+\vec \grad{D} 
\end{equation}
where, $H_{1}$ \& $H_{2}$ are Nambu Hamiltonians that are derived from $\vec v_{ND}$-part and $D$ is the dissipation function, which is derived from $\vec v_{D}$-part.

In many linear physical systems, we find that a gradient dissipative(dissipative force) term is added to the conservative Hamiltonian. Therefore, this kind of decomposition in chaotic systems is in line with the standard equation of motion of a large class of known linear systems. It is important to note that this decomposition is not unique. A different choices can be made for the $\vec v_{ND}$ and $\vec v_{D}$ parts, while maintaining the Helmholtz-Hodge decomposition. Let us illustrate this decomposition using the Lorenz and Rössler systems as examples.

The Lorenz system is given by the equations of motion $v =(\sigma (y-x),\;x(r-z)-y,\; xy-bz)$.  We can have following pair of decompositions: $\vec v_{ND}=(\sigma y,\ x(r-z), xy)$ while $\vec v_{D}=(-\sigma x, -y, -bz)$; or $\vec v_{ND}=(-ry+\sigma y, -xz, xy)$ while $\vec v_{D}=(-\sigma x+ry, -y+rx, -bz)$.  Both the cases satisfy the required conditions: $\vec\grad\cdot \vec v_{ND}=0$ and $\vec\grad\cross \vec v_{D}=0$.  Additionally,  $\vec v_{ND}=(\vec\grad{H_{1}}\cross \vec\grad{H_{2}})$  indicates that $H_{1}$ \& $H_{2} $  are Nambu Hamiltonians that are derived from the non-dissipative component. Thus, the choice of decomposition in the Lorenz system is not unique.
 
Another example is the Rössler system, which is defined by the equations of motion as $v= (-y-z ,\;x+ay,\;b+z(x-c))$. The two different possible decompositions are : $\vec v_{ND}=(-y-z-\frac{z^2}{2}, x, b)$, while $\vec v_{D}=(\frac{z^2}{2}, ay, z(x-c))$ or $\vec v_{ND}=(-y-z-\frac{x^2}{2}, x, b+xz)$, while $\vec v_{D}=(\frac{x^2}{2}, ay, -cz)$. It is not necessary that the decomposition parts contain only the terms of the vector field. Sometime, to achive physically realizable decomposition, one may add and subtract terms in the vector field,  as seen in the case of the Rössler attractor \cite{Axenides_2010}. 

When a system exhibits multiple valid decompositions that satisfy all the key conditions outlined earlier (i.e. $\vec\grad\cdot \vec v_{ND}=0$, $\vec\grad\cross \vec v_{D}=0$, $\vec v_{ND}=(\vec\grad{H_{1}}\cross \vec\grad{H_{2}})$ \& $\vec v_{D}=\vec\grad{D}$), the resulting Nambu Hamiltonian surfaces in the phase space exhibit similar nature. However, these Hamiltonians can differ in their mathematical expressions and may possess different orientations depending on the chosen parameter values. We  illustrate this point using the Lorenz system (details are available in Appendix \ref{BB}). This leads us to a compelling conclusion: the derived Nambu Hamiltonian functions stand as uniquely defined surfaces for a specific system, though the decomposition of the corresponding vector field is not unique. 
\section{\label{sec3}The four-wing system}
The dynamical equations for a four-wing chaotic attractor are expressed as follows:
\begin{equation}\label{eq1}
	\begin{cases}
		\dot{x} =ax+cyz \\
		\dot{y}=bx+dy-xz \\
		\dot{z}=ez+fxy 
	\end{cases}\,.
\end{equation}

where ($a$, $b$, $c$, $d$, $e$, $f$) are system parameters and $(x, y, z)$ are system variables. The system exhibits multi-lobe four-wing chaotic dynamics for specific parameter values: $a=0.2$, $b=-0.01$, $c=1$, $d=-0.4$, $e=-1$, $f=-1$. This particular system, with these fixed parameter values, was first analyzed by Z. Wang \textit{et al.} \cite{wang20093}. The summery of the fundamental properties of the system (\ref{eq1}) as reported in \cite{wang20093} and from our further analysis are mentioned below:\\

(a)	\textit{Dissipativity}: 
	The divergence of this four-wing system is found to be
	\begin{equation}
		\vec\grad\cdot {\vec v} =\frac{\partial \dot{x}}{\partial x}+ \frac{\partial \dot{y}}{\partial y}+\frac{\partial \dot{z}}{\partial z} =(a+d+e)=-1.2 <0
	\end{equation}
	
	This shows the four wing system is dissipative.\\

(b)	\textit{Equilibrium points}:	
	 The system exhibits five equilibrium points, which can be identified by solving the equations ($\dot{x}=0$, $\dot{y}=0$, $\dot{z}=0$). The stability of the equlibrium points are found by the eigenvalues of the Jacobian matrix (\textbf{J}), which is given by
	\begin{equation}{\label{jac}}
		\textbf{J}=\mqty(a & cz_0 & cy_0 \\ (b-z_0) & d & -x_0 \\ fy_0 & fx_0 & e ) 
	\end{equation}
where $x_0$, $y_0$ and $z_0$ are the equilibrium points. All the equilibrium points are reported in \cite{wang20093} and are also verified by our calculation.
	
\begin{figure}[htbp!]
		\centering
		{\includegraphics[width=1.0\linewidth]{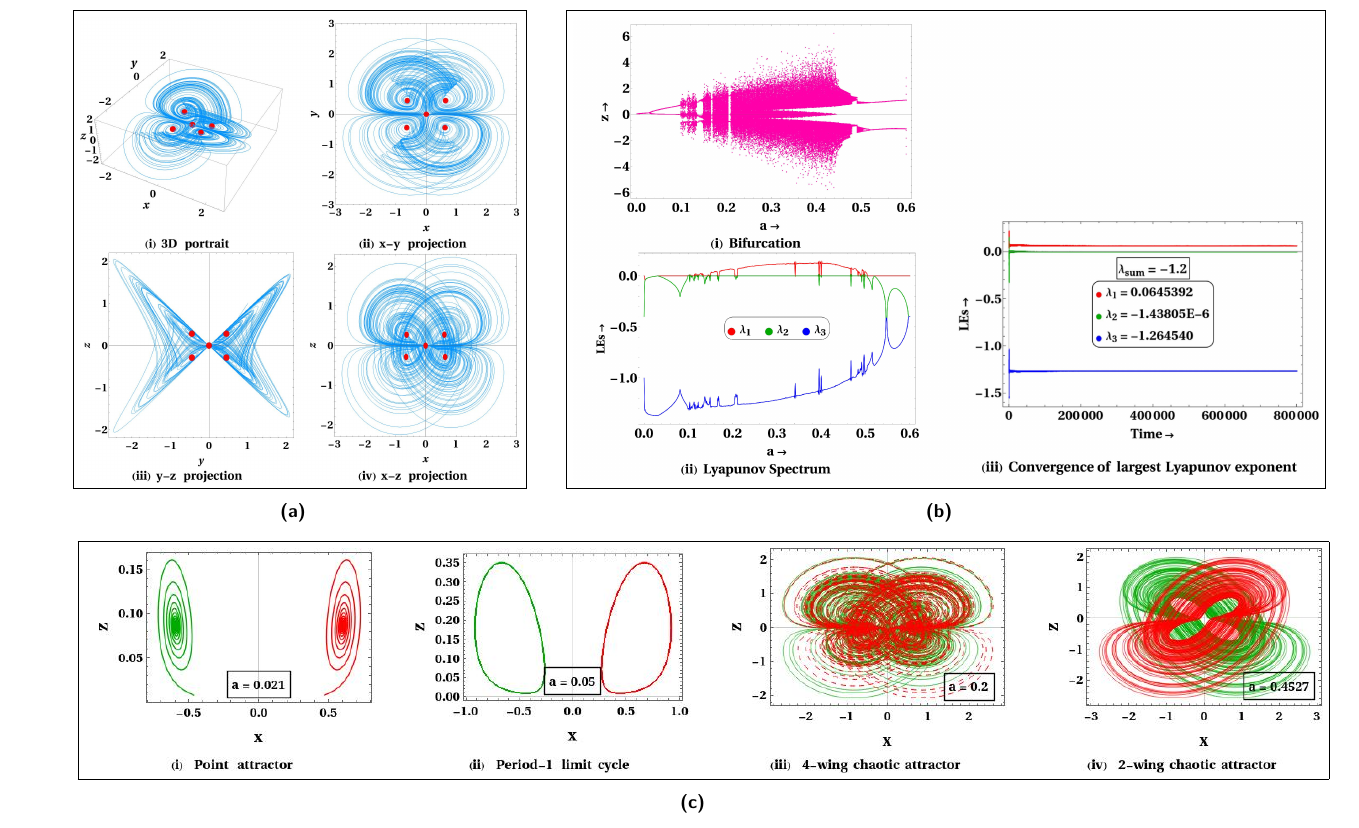}}
		\caption{The sub-figures a(i-iv) illustrate the phase-space trajectories of the system (\ref{eq1}) and the red colour points represent its equilibrium points. The sub-figures b(i) \& b(ii) indicate the bifurcation diagram and Lyapunov spectrum of the system for a varying parameter `a', respectively; whereas b(iii) confirms the convergence of the largest Lyapunov exponent. The system (\ref{eq1}) shows bi-stability for different values of parameter `a' as depicted in sub-figures c(i-iv) and respective initial conditions for green and red colour orbits are (x(0), y(0), z(0))=($\pm 1$, $\pm 1$, 1).}
		\label{ps}
		\end{figure}

	From FIG. \ref{ps}(a) we observe that, with the exception of the central equilibrium point at the origin, all other four equilibrium points (represented by the red dots) are associated with a specific wing of the multi-lobe system for a selected set of parameters.  The eigen-values mentioned in TABLE \ref{table1} show that, out of the five equilibrium points, four are index-2 saddle focus points, while the remaining central equilibrium point at $(x, y, z) = (0, 0, 0)$ is an index-1 saddle focus point. An index-2 saddle focus point is characterized by one negative real eigenvalue and two complex conjugate eigenvalues with positive real parts. These index-2 saddle focus points act as dynamic centres for the `wings' and these equilibrium points are responsible for generating the four-wing chaotic attractor \cite{8866023, Jiang2020, zhang2013constructing}. \\
	
	\begin{table}[h]
		\centering
		\footnotesize
		\caption{\label{table1} The parameter values for the four-wing system, their five equilibria, and their corresponding three eigenvalues of the Jacobian matrix at each equilibrium point.}
		
		\addtolength{\tabcolsep}{-1.3pt}
		\renewcommand{\arraystretch}{1}
		\begin{tabular}{c l c l}
			\toprule[0.07cm]
			\bf{Parameters: $(a, b, c, d, e, f)$}& \bf{Equilibrium points: $(x, y, z)$} & \bf{Eigen values of Jacobian: $\lambda_1, \lambda_2, \lambda_3$}\\    
			
			\midrule
			
			\multirow{1}{*}{
				$(0.2, -0.01, 1, -0.4, -1, -1)$}& 
			
			$\begin{cases}
				(0,0,0) \\
				(-0.643735, -0.447214, -0.287887)\\
				(-0.621374, 0.447214, 0.277887)\\
				(0.621374, -0.447214, 0.277887)\\
				(0.643735, 0.447214, -0.287887)
			\end{cases}$& 
			$\begin{cases} 
				-1, -0.4, 0.2 \\
				-1.38118 , 0.0905921 \pm 0.477124 i \\
				-1.35983 , 0.0799151 \pm 0.474184 i\\
				-1.35983 , 0.0799151 \pm 0.474184 i\\
				-1.38118 , 0.0905921 \pm 0.477124 i 
			\end{cases}$
			
			\\ \addlinespace
			
			\bottomrule[0.07cm]
		\end{tabular}
	\end{table}
	

(c)	\textit{Bifurcation diagram and Lyapunov Exponents} : 	
	The dynamical behaviour of the system (\ref{eq1}) is analyzed by solving the dynamical equations numerically using the standard fourth-order Runge-Kutta algorithm and the Lyapunov exponent spectrum is computed using the Gram-Schmidt orthonormalization method \cite{christiansen1997computing}. For our calculation, we consider the same set of parameters as reported in \cite{wang20093} i.e. $(a, b, c, d, e, f)$ = ($0.2$, $-0.01$, $1$, $-0.4$, $-1$, $-1$) with the initial conditions $(1,1,1)$.  Various numerically found  phase portraits are illustrated in FIG.\ref{ps}(a). Points with red colour represent its equilibrium points. Instead of calculating bifurcation and Lyapunov exponent diagrams by varying `b' as reported in \cite{wang20093}, we calculate the same by varying the parameter `a' within a range of ($0 < a < 0.6$) while keeping rest of the parameters fixed.  The reason for doing so is that the two Nambu surfaces (Nambu doublet)  depend on `b' whereas the dissipative part depends on `a'. From the expression of the doublet (equation(\ref{hh})), it is clear that the nature of the surfaces does not change if `b' is varied, while a small change in `a' affects the condition for localization, equation (\ref{Ssdot}), due to  change in dissipative part. Therefore, it is important to know the region of chaos in the `a' parameter space. The sub-figures shown in FIG.\ref{ps}(a)  confirm that the system (\ref{eq1}) has extremely rich dynamics. The bifurcation diagram (see FIG.\ref{ps}b(i)) and  the corresponding Lyapunov spectrum (largest positive Lyapunov exponent) (see FIG.\ref{ps}b(ii)) indicate
	the existence of  chaotic regimes  for the variation of control parameter `a'. The three Lyapunov exponents of the system are found to be $\lambda_{1}$=0.0645392, $\lambda_{2}$=-1.43805E-6, $\lambda_{3}$=-1.26454 and the sum of all three $\lambda_{i}$'s is determined to be negative i.e. -1.2, which confirms that the system is a dissipative chaotic one. We checked the convergence of the calculation of Lyapunov exponent (see FIG.\ref{ps}b(iii)) for numerical accuracy.\\

(d)	\textit{Fractal dimension or Kaplan-Yorke dimension ($D_{KY}$)}:	
	 A strange attractor's fractal dimension, which has been regarded as its most fundamental characteristic, is a measurement of its geometric scaling properties or its `complexity'. Characterization of the fractional dimension of the strange attractors generated by chaotic flows are approached in a variety of ways. These techniques can be divided into two groups i.e. topology-based and dynamics-based methods \cite{clark1990estimating, datseris2023estimating}. 
	 
	 The Kaplan-Yorke Dimension ($D_{KY}$) \cite{bakri2025note} is a measure of the fractal dimension of a chaotic attractor derived from its Lyapunov exponents. For a given dynamical system having n no. of Lyapunov exponents $\lambda_{1} \geq \lambda_{2} \geq \lambda_{3}...\geq \lambda_{n}$, the $D_{KY}$ is defined as 
	 \begin{equation}
	 	\label{eq2}
	 	D_{KY} = m + \frac{\sum_{i=1}^{m} \lambda_{i}} {\abs{\lambda_{m+1}}}
	 \end{equation}
where  m = the largest integer such that $\sum_{i=1}^{m} \lambda_{i} \geq 0$ \& $\sum_{i=1}^{m+1} \lambda_{i}= -ve $ and 
	 $\lambda_{m+1}$ = first Lyapunov exponent for which  $\sum \lambda_{i}$ becomes negative.
	 
For the present system, we find the fractal dimension as follows:
\begin{equation}
	  D_{KY} = 2 + \frac{0.0645377} {\abs{-1.26454}}= 2 + 0.0510365 \approx 2.05104
\end{equation}
	  
 The non-integer $D_{KY}$ confirms the attractor is a  fractal geometry. The value suggests the structure of attractor is more complex than a simple surface but not fully volume-filling. We notice that $D_{KY}$ lies between 2 and 3, indicating the four-wing chaotic attractor as a `folded' or `layered' structure. The exact value of $D_{KY}$ depends on the system parameters, because parameters play crucial role in determining the chaotic dynamics. Generally, increasing the number of wings or lobes in multi-wing attractors increases $D_{KY}$ due to enhanced geometrical and dynamical complexity. But, the exact change depends on how the Lyapunov exponents shift. \\

(e)	 \textit{Bi-stability}: 	
	We observe that the system (\ref{eq1}) is invariant under the change of coordinates: (x, y, z) → (-x, -y, z) for a fixed set of parameters, demonstrating the existence of rotational symmetry with respect to the z-axis. Thus,  coexisting symmetric attractors are expected. Though multi-stability is not a common and fundamental property of all chaotic attractors, it is still an exciting phenomenon of a non-linear chaotic system. The phase space trajectories shown in FIG.\ref{ps}(c) indicate the bi-stability of the system (\ref{eq1}) and it is obtained  by changing the initial conditions. We observe coexisting symmetric attractors for different values of the parameter `a', such as point attractor for $a = 0.021$, period-1 limit cycle for $a = 0.05$, four-wing chaotic attractor for $a = 0.2$, and two-wing chaotic attractor for $a = 0.4527$ as depicted in FIG.\ref{ps}c(i-iv). Respective initial conditions for green and red colour orbits are considered to be (x(0), y(0), z(0))=($\pm 1$, $\pm 1$, $1$).
	
\section{\label{sec4} Analysis of four-wing attractor based on Nambu mechanics}
As stated in the introduction that the primary objective of the present research is to explore the mechanism of formation of four wing lobes geometry of the attractor in the phase space and to find out the condition for its localization in a finite region in the phase space.

The first step towards this is to decompose the velocity vector flow into two distinct components: a conservative  non-dissipative part ($\vec v_{ND}$) and the other a non-conservative  dissipative part ($\vec v_{ND}$) in such a way that $v=\vec v_{ND}+\vec v_{D}$, while satisfying all the key conditions discussed in the section \ref{sec2}.\\

The valid decompositions are given by:
\begin{equation}\label{nd}
	\text{Non-dissipative part:}\quad
	\vec v_{ND}=(cyz, bx-xz, fxy)
\end{equation}
\begin{equation}\label{d}
	\text{Dissipative part:}\quad
	\vec v_{D}=(ax, dy, ez)
\end{equation}
\begin{equation}\label{part2}
	\text{where}\quad
	\vec v_{D}=\vec\grad{D}
\end{equation}
\subsection{\label{sec4a} Analysis of four-wing non-dissipative part}
The non-dissipative part can be written in the form of 

\begin{equation}\label{part1}
	\vec v_{ND}=\vec\grad{H_{1}}\cross\vec\grad{H_{2}}
\end{equation}

The two Nambu Hamiltonians $H_{1}$ \& $H_{2}$ for this system are (see Appendix \ref{CC} for detailed derivation):
\begin{equation}\label{hh}
	H_{1}=\frac{1}{2}x^{2}-\frac{1}{2}\frac{c}{f}z^{2}\quad,\quad   
	H_{2}=\frac{1}{2}fy^{2}+\frac{1}{2}z^{2}-bz
\end{equation} 

From equation (\ref{hh}) we observe that the Nambu doublet $(H_1, H_2)$ depend on the parameters $b, c, f$. The values of these parameters are so chosen that the system is in chaotic regime.  As discussed in the section \ref{sec2} and the equation (\ref{NDeq}), the trajectory lies on the intersection of the two surfaces given by $H_1 =$ constant and $H_2 =$ constant respectively. These constants are $H_1(t_0)$ and $H_2(t_0)$ respectively, which are determined from the initial condition at time $t_0$ . Therefore, nature of a trajectory depends on the nature of the above Hamiltonian(energy) surfaces. The expressions for the two Hamiltonians show that the nature of the surfaces depend on the sign of the parameters $\frac{c}{f}$ and $f$. If both the parameters have positive sign, the first surface is hyperbolic, where as the other one is cylindrical. If they have negative sign, nature of surfaces get interchanged. On the other hand, if they have opposite sign, both the surfaces are either cylindrical or hyperbolic. This clearly explains why specific geometry of trajectories depends on these parameters.

Since $H_1$ and $H_2$ are obtained from the non-dissipative part, we expect $\dot H_{1}= 0$ and $\dot H_{2}= 0$. This is indeed true (it is verified from the equation (\ref{nondisSdot}) as $\dot S_{ND}= 0$). Thus, $H_1$ and $H_2$ are conserved quantities in the absence of dissipative part, $\vec v_{D}(=\vec \grad D$), where $D$ is the dissipation function. These Hamiltonian surfaces are two-dimensional surfaces in a three-dimensional space, and their intersection forms a one-dimensional curve. This property makes Nambu mechanics a geometrically intuitive framework, where conservation laws and permissible trajectories are unified in a single mathematical structure or expression (i.e. $\vec\grad{H_{1}}\cross\vec\grad{H_{2}}$). As a result, the cross product is not merely a convenient choice; it is indeed the only vector that points along the intersection of the constrained surfaces $H_1$ \& $H_2$.\\

In a nutshell, we can say that the non-dissipative trajectory is obtained from the intersection of the two surfaces defined by 
\begin{equation}\label{sur}
	H_1 = H_1(t_0) \quad \& \quad
	H_2 = H_2(t_0)
\end{equation}
where $H_1(t_0)$ and $H_2(t_0)$ are the initial values at  $(x_0, y_0, z_0)$.

\begin{figure}[htbp!]
	\centering
	{\includegraphics[width=0.7\linewidth]{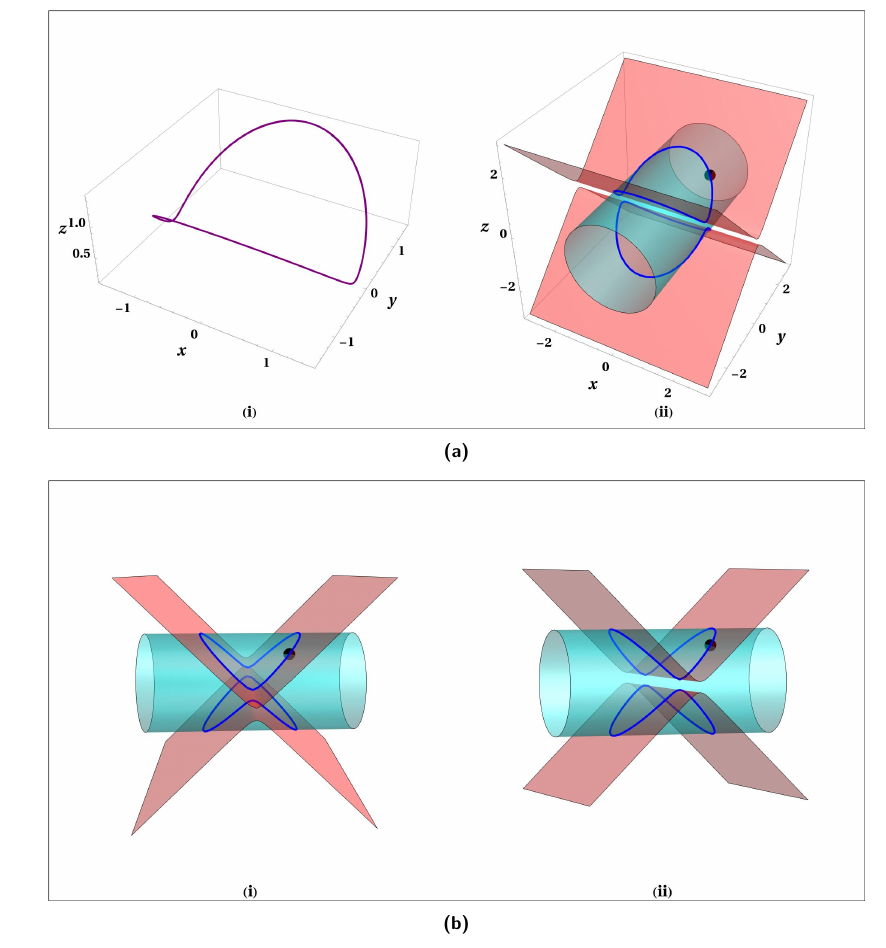}}
	
		\caption{The sub-figures a(i) \& (ii) represent the non-dissipative orbit of system (\ref{eq1}) and  the intersecting orbit (blue colour) of it's Nambu surfaces $H_1=H_1(t_0)$ \& $H_2=H_2(t_0)$ at the initial condition $(x_0, y_0, z_0) =(1, 1, 1)$, respectively. The sub-figures b(i) \& (ii) illustrate four lobe geometry of intersecting orbits of Nambu doublets from two different viewpoints (or) orientations of these Nambu surfaces. The black colour dots shown in the sub-figure a(i)\& b(i-ii) indicate the position of the particle at the same initial condition $(1,1,1)$.}
		\label{intersect}
\end{figure}

The two Nambu surfaces and their intersecting orbit for the chosen parameter values, $b=-0.01, c=1, f=-1$ are illustrated in  FIG. \ref{intersect}a(ii). For comparison purpose, we also plotted the non-dissipative orbit from direct numerical solution to the equation (\ref{eq1}), shown in FIG. \ref{intersect}a(i).

 The $H_{1}$ surface represents a cylinder oriented along the y-axis, while the $H_{2}$ surface corresponds to a hyperboloid, oriented along the x-axis as shown in  FIG. \ref{intersect}. This is expected for the chosen sign of the parameters.  FIG. \ref{intersect}a(ii) clearly shows that both the intersecting orbits (blue colour orbit) are identical to the actual non-dissipative orbit of the system (shown in FIG. \ref{intersect}a(i)) obtained from direct numerical solution and this validates the theory of Nambu formalism. Therefore, in the absence of the dissipation, one of the two intersecting orbits is the actual trajectory depending on the initial condition, corresponds to the figure FIG. \ref{intersect}a(i). Each intersecting orbit corresponds two lobes or wings of the full four wing attractor when dissipation is added. This will be discussed in the subsequent sections. The orbit is identified as a homoclinic limit cycle. Sub-figures of FIG. \ref{intersect}(b) show the intersection of these two Nambu surfaces when viewed from two different angles/orientations, illustrating three dimensional geometry of the desired four-lobed structure.  
 
The sign of the parameters $\frac{c}{f}$ and $f$ determines the geometry of the surfaces, but the magnitudes  and initial conditions determine the sizes and their relative positions. This means intersection of the surfaces depends on these two factors and therefore, the actual trajectory depends on these parameters and the initial conditions (justifying the presence of basin of attraction).

 In the case of Lorenz and Rössler attractors, we observe that there may be several possible decompositions (refer to subsection \ref{sec2B}). This is also true in the case of the four-wing attractor. Another decomposition possibility for four-wing attractor other than the one discussed above is $\vec v_{ND}=(cyz-by, -xz, fxy) \quad \& \quad  \vec v_{D}=(ax+by, bx+dy, ez)$. Like Lorenz and Rössler system, this four-wing system also produces a unique and identical Nambu doublet (see Appendix \ref{DD} for the detailed derivation) irrespective of  different decomposition choices, indicating the uniqueness of Nambu Hamiltonian functions for a specific system. 
\subsection{\label{sec4b}Different Nambu surfaces}
As discussed in the introduction, there was an attempt to establish localization of Lorenz attractor by introducing several geometric surfaces \cite{doi:10.1080/02681119508806207}. The same surfaces  are also generated from the parent Nambu doublet of Lorenz attractor to discuss localization of the attractor without involving the intersection of two surfaces \cite{Roupas_2012}. The later approach is based on the transformation property of the doublet as discussed in the section \ref{sec2}. The same is true for the present four-wing system also. The Nambu doublet $h=(H_{1},H_{2})$ can be transformed to generate new transformed Nambu surfaces with the help of a transformation matrix, whose determinant should be equal to one. This means the transformation is canonical. Therefore, transformed doublet can be written as
\begin{equation}
	h'=A h 	\Rightarrow \mqty(H_{1}'\\ H_{2}')
	=\mqty(\alpha & \beta \\ \gamma & \delta) \mqty(H_{1} \\ H_{2})
\end{equation}

The transformation matrix \textbf{$A$} has a general form of  A = $\mqty(\alpha & \beta \\ \gamma & \delta)$, $\forall \;\; (\alpha, \beta,\gamma,\delta) \in R $ such that the determinant of transformation matrix  \textbf{$A$} \; i.e. $(\alpha\delta-\beta\gamma) =1 $. By various choices of $\alpha$, $\beta$,$\gamma$, $\delta$ with the restriction $(\alpha\delta-\beta\gamma) =1 $, we can construct different sets of new transformed Nambu doublets. With a specific choice of \textbf{$A$}, the new transformed doublet is nothing but the interchange between $H_1$ and $H_2$. Therefore, it is possible to write a single expression for the both the surfaces. Let us represent these two  Nambu surfaces by a generalized expression for single surface  $S$, involving $\alpha$, $\beta$,$\gamma$, and $\delta$ as
\begin{equation*}                                                         
\mqty(H_{1}'\\ H_{2}')
=\mqty(\alpha & \beta \\ \gamma & \delta) \mqty(\frac{1}{2}x^{2}-\frac{1}{2}\frac{c}{f}z^{2} \\ \frac{1}{2}fy^{2}+\frac{1}{2}z^{2}-bz)= \mqty(S(x, y, z;\alpha,\beta)\\  S(x, y, z;\gamma,\delta))
\end{equation*}
where, $S(x, y, z;\alpha,\beta)$ corresponds to a quadratic surface $H_{1}'$, while replacing ($\alpha$ \& $\beta$) with ($\gamma$ \& $\delta$) respectively it gives $H_{2}'$ surface. However, $S(x, y, z;\alpha,\beta)$ \& $S(x, y, z;\gamma,\delta)$ has the same expression in terms of $(\alpha,\beta)$ or $(\gamma,\delta)$ respectively.\\
Thus, for the four-wing system (\ref{eq1}), the equation (\ref{hh}) can be casted into the form
\begin{equation*}
 S(x, y, z;\alpha,\beta)= \alpha H_{1}+\beta H_{2}=\alpha \left(\frac{1}{2}x^{2}-\frac{1}{2}\frac{c}{f}z^{2}\right)+\beta \left(\frac{1}{2}fy^{2}+\frac{1}{2}z^{2}-bz\right)
\end{equation*}
\begin{equation}\label{S1}
	\Rightarrow  S(x, y, z;\alpha,\beta)=\frac{1}{2}\alpha x ^{2}+\frac{1}{2}\beta fy^ {2}+ \frac{1}{2}\left(\beta-\alpha\frac{c}{f}\right) z ^{2}- \beta bz 
\end{equation}
For $\alpha = 1$ \& $\beta = 0$, equation (\ref{S1}) yields $H_{1}$ surface; whereas for $\alpha = 0$ \& $\beta = 1$, it yields  $H_{2}$ surface.

\begin{figure}[htbp!]
	\centering
	{\includegraphics[width=0.88\linewidth]{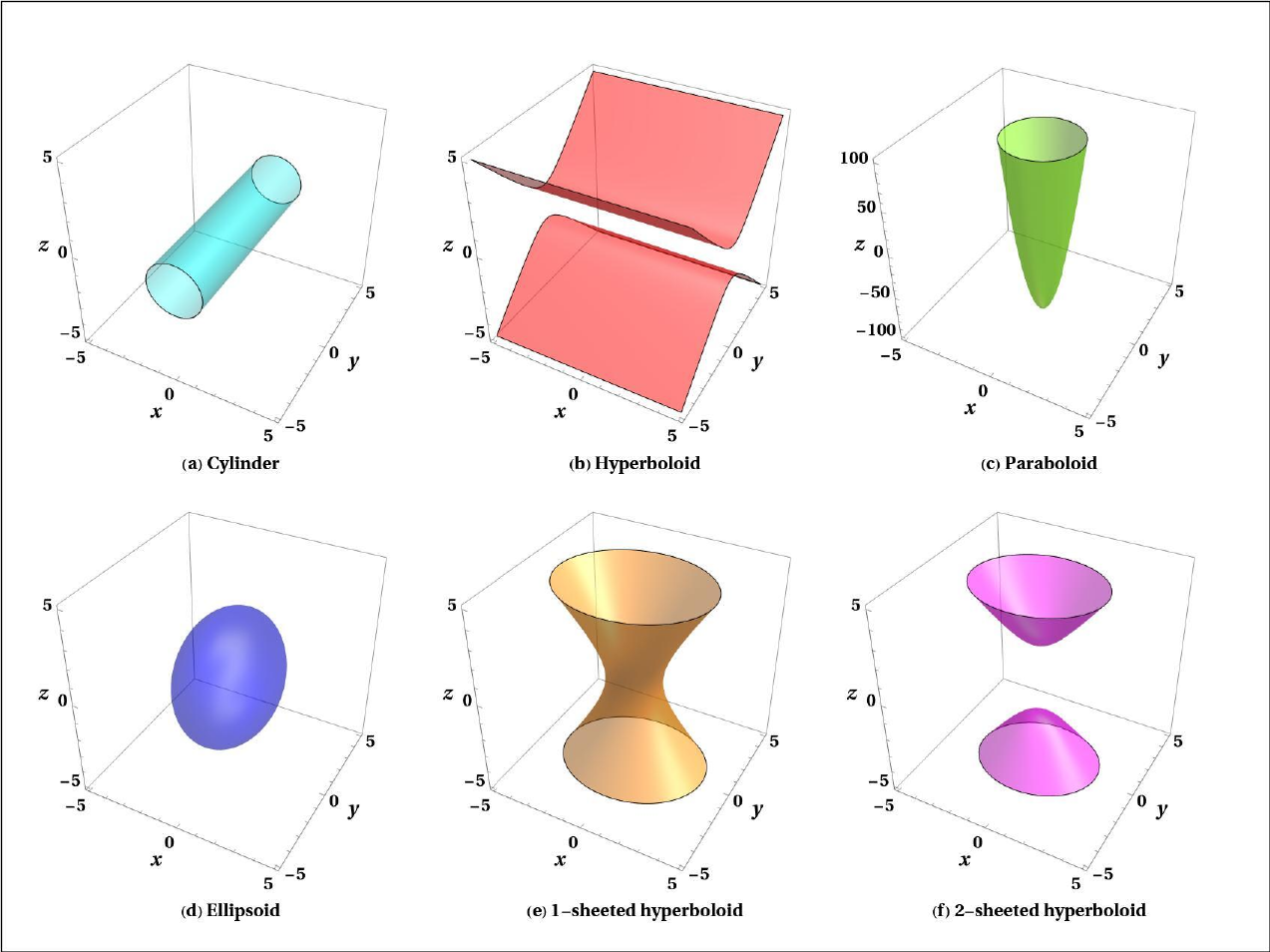}}
	
	\caption{Different possible transformed Nambu surfaces that can be generated from equation (\ref{S}) for different $\eta$ values.}
	\label{transSur}
\end{figure}

%
%
%
%
\begin{figure}[htbp!]
	\centering
	{\includegraphics[width=0.88\linewidth]{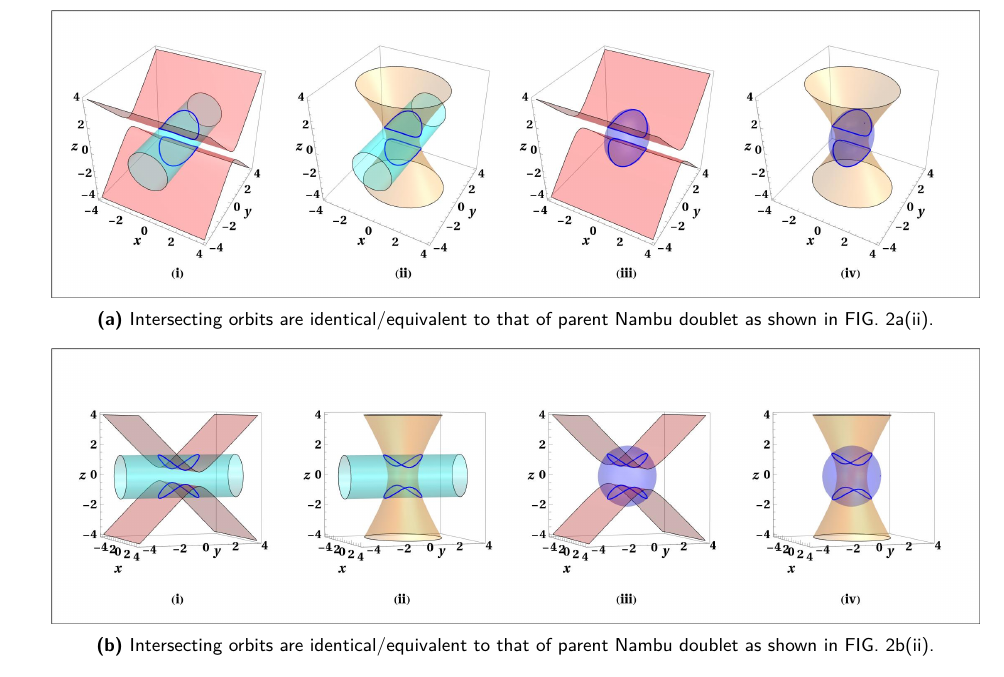}}
	
\caption{ In all the sub-figures, the blue colour orbit represents intersection of different transformed Nambu doublets and the intersecting orbits are found to be identical/equivalent to that of parent Nambu doublet shown in FIG. \ref{intersect}a(ii) \& b(ii), indicating the canonical transformation of Nambu functions.}
	\label{identicalIOs}
\end{figure}

With rescaling, the above equation (\ref{S1}) can be rewritten in terms of a single parameter $\eta$, which is given by
\begin{equation}\label{S}
	\begin{aligned}
	S(x,y,z;\eta) =\frac{1}{\alpha} * S(x, y, z;\alpha,\beta) = \frac{1}{2}x ^{2}+\frac{1}{2}\eta fy^ {2}+ \frac{1}{2}\left(\eta-\frac{c}{f}\right) z ^{2}- \eta bz\\
    S(x,y,z;\eta)=\frac{1}{2}x ^{2}+\frac{1}{2}\eta fy^ {2}+ \frac{1}{2}\left(\eta-\frac{c}{f}\right)\left(\left( z-\rho\right) ^2 - {\rho}^2\right)
	\end{aligned}
\end{equation}
where, $\eta =  \frac{\beta}{\alpha}$  and  $\rho =\frac{b\eta}{\left(\eta-\frac{c}{f}\right)}$. The last equation is not valid for $\eta = \frac{c}{f}$.
It is clear that for $\alpha = 1$ \& $\beta = 0$ $\Rightarrow \eta = 0$ and this gives the surface $H_{1}$ of equation (\ref{hh}), while for $\alpha = 0$ \& $\beta = 1$ $\Rightarrow \eta = \infty$ and this gives the surface $H_{2}$ of equation (\ref{hh}). We can derive other possible transformed Nambu surfaces from the equation (\ref{S}) depending on various choices of $\eta$ value. Since all these transformed surfaces are generated with the help of a canonical transformation, so these surfaces are not qualitatively different. They fall under only four types of surfaces, namely
\begin{enumerate}
	
\item {\bf{Cylindrical Surface:}} For $\eta = 0$ , the equation (\ref{S}) reduces to 
\begin{equation}\label{Scy}
	S(x,y,z;\eta) =  \frac{1}{2}x ^{2} - \frac{1}{2}\frac{c}{f}z^2 
\end{equation}
  and for $\frac{c}{f} < 0$, the above equation represents a cylinder with its axis as y-axis.
 
\item {\bf{Paraboloid:}} For $\eta = \frac{c}{f}$, the equation (\ref{S}) reduces to
\begin{equation}\label{Spa}
	S(x,y,z;\eta) =  \frac{1}{2}x ^{2}+\frac{1}{2} c y^ {2} -\frac{c}{f} b z
\end{equation} 
This equation represents a paraboloid, oriented along the z-axis provided $c$ is positive.

\item {\bf{Ellipsoid :}} For $\eta > \frac{c}{f}$, the equation (\ref{S}) represents an ellipsoid  with centre at ($0,0,\rho$) if $\eta{f} >0$. Equation (\ref{S}) then takes the form
\begin{equation}\label{Sel}
		S(x,y,z;\eta)+ \frac{1}{2}\left(\eta-\frac{c}{f}\right){\rho}^2=\frac{1}{2}x ^{2}+\frac{1}{2}\eta fy^ {2}+ \frac{1}{2}\left(\eta-\frac{c}{f}\right)\left( z-\rho\right) ^2 
\end{equation}

\item {\bf{Hyperboloid :}} For $\eta = 0$, the surface is though represented by the equation (\ref{Scy}), but it is a hyperboloid for $\frac{c}{f}>0$. For $\eta = \frac{c}{f}$, the equation is same as the equation (\ref{Spa}) but it is a hyperboloid if $c<0$. For $\eta > \frac{c}{f}$, the surface (\ref{S}) is a hyperbola for $\eta{f} <0$ and is given by the equation (\ref{Sel}). Finally for  $\eta < \frac{c}{f}$, it is a hyperboloid for all possible values of $c$ and $f$ because $z^2$ terms remains negative and its equation is again given by (\ref{Sel}). The hyperboloid is one  or two sheeted depending on the sign of the left hand side of the equation (\ref{Sel}).

\end{enumerate}

FIG. \ref{transSur} represents the  geometrical shapes of all possible transformed Nambu surfaces. It is to be noted that if one surface, $H_1 = H_1(0)$ is a hyperboloid,  the other $H_2 = H_2(0)$ is necessarily be cylindrical or ellipsoid or paraboloid and vice versa, so that we can get four-lobed intersecting orbit. The intersection of any two transformed Nambu surfaces are shown in FIG. \ref{identicalIOs} . Here, it is observed that the shape of intersecting orbits in all cases
are found to be the similar/identical, which is equivalent to the non-dissipative orbit of the four-wing
system (see FIG. \ref{intersect}a(i)), depicting the canonical transformation of Nambu functions. The only
difference between the intersecting orbits depicted in FIG. \ref{identicalIOs}(a) \& (b) is that these are plotted for
two different orientations of concerned Nambu surfaces. Moreover, it is noticed that orbits shown in
FIG. \ref{identicalIOs}(a) \& (b) are similar to the orbits illustrated in FIG. \ref{intersect}a(ii) \& FIG. \ref{intersect}b(ii) respectively.  This is expected for an unique dynamics. 

\subsection{\label{sec4c}Effect of the dissipative part and formation of complete chaotic orbits}

 We know that when dissipation comes into picture, the dynamics of the system continuously change over time, indicating the variation or continuos deformation of the Nambu surfaces. It is verified by calculating time derivation of Nambu surfaces.
 
The time derivative of the generalized surface $S$, using the equations (\ref{nd}) \& (\ref{d}), is given by
\begin{equation}\label{Ssdot}
	\dot S= \vec\grad S \cdot \vec v=\vec\grad S \cdot (\vec v_{ND}+\vec v_{D})=\vec\grad S \cdot \vec v_{ND} + \vec\grad S \cdot \vec v_{D}=\dot S_{ND}+\dot S_{D}
\end{equation}
The first term is zero as expected i.e.
\begin{equation}\label{nondisSdot}
\dot S_{ND}=\vec \grad S \cdot \vec v_{ND} = \qty\bigg(x,\; \eta fy,\; \left(\eta-\frac{c}{f}\right) (z-\rho)) \cdot \qty\bigg(cyz,\; bx-xz,\; fxy)=0
\end{equation}
The second term contributes only to the time evolution of the surface, because
\begin{equation*}
\dot S_{D}=  \vec\grad S \cdot \vec v_{D}=\qty\bigg(x,\; \eta fy,\; \left(\eta-\frac{c}{f}\right) (z-\rho)) \cdot \qty\bigg(ax,\;  dy,\;  ez)
\end{equation*}
\begin{equation}\label{disSdot}
	\Rightarrow \dot S_{D}=\vec\grad S \cdot \vec v_{D}= ax ^{2}+\eta fdy^ {2}+ e\left(\eta-\frac{c}{f}\right)z(z-\rho)  \neq 0
\end{equation}
Using equations (\ref{nondisSdot}) and (\ref{disSdot}), we can rewrite the equation (\ref{Ssdot}) as 
\begin{equation}
\dot S=	\dot S_{ND}+\dot S_{D}=0+\dot S_{D}\Rightarrow \dot S=\dot S_{D}
\end{equation}
Therefore, we can say that the surface $S$ evolves according to
\begin{equation}\label{DotS}
	\begin{aligned}
\dot S = ax ^{2}+\eta fdy^ {2}+ e\left(\eta-\frac{c}{f}\right)z^2 - \eta{ebz}\\	
	\dot S = ax ^{2}+\eta fdy^ {2}+ e\left(\eta-\frac{c}{f}\right)\left(\left( z-\frac{\rho}{2}\right) ^2 - \left(\frac{\rho}{2}\right)^2\right)
\end{aligned}
\end{equation}
In the present case the dissipative term $\vec v_D$ is linear in the state variables. Hence, the full dynamical equation (\ref{eq1}) can be rewritten into the form of equation (\ref{NDeq}) with time dependent Hamiltonians, which is given by :

\begin{equation}
	\label{DE}
	\begin{aligned}
		\dot{x}=\left(\vec\grad{H_{1}}\cross\vec\grad{H_{2}}\right)_{1} -{\eta}_1 x\\
		\dot{y}=\left(\vec\grad{H_{1}}\cross\vec\grad{H_{2}}\right)_{2} -{\eta}_2 y\\
		\dot{z}=\left(\vec\grad{H_{1}}\cross\vec\grad{H_{2}}\right)_{3} -{\eta}_3 z\\
	\end{aligned}
\end{equation}
where  $\eta_1 = -a$, $\eta_2 = -d$ \& $\eta_3 = -e$ for the present four-wing system. Let us define, a set of new variables $u$, $v$, $w$ and new time $\tau$. These new variables are related to the old variables $x, y, z$ and old time t as : 
\begin{equation}
 u = e^{\eta_1 t}x,\;  v = e^{\eta_2 t}y,\;  w = e^{\eta_3 t}z , \; \tau = \frac{1}{(\eta_1 + \eta_2+ \eta_3)}e^{\left(\eta_1+ \eta_2 + \eta_3\right)t}
\end{equation}
In terms of new variables ($u, v, w$), the dynamical system becomes (detailed derivation is given in the Appendix \ref{EE})\\
\begin{equation}
	\label{eq2}
	\frac{d}{d\tau}(u,v,w)= \vec\grad_{\vec q} {H_{1}}\cross\vec\grad_{\vec q}{H_{2}}
\end{equation}

where $\vec q \equiv (u,v,w)$ and Hamiltonians are time dependent i.e. $H_{i} = H_{i}\left(e^{-\eta_1 t}u, e^{-\eta_2 t}v, e^{-\eta_3t}w\right)$ , i = 1,2 .

\begin{figure}[htbp!]
			\centering
			{\includegraphics[width=1\linewidth]{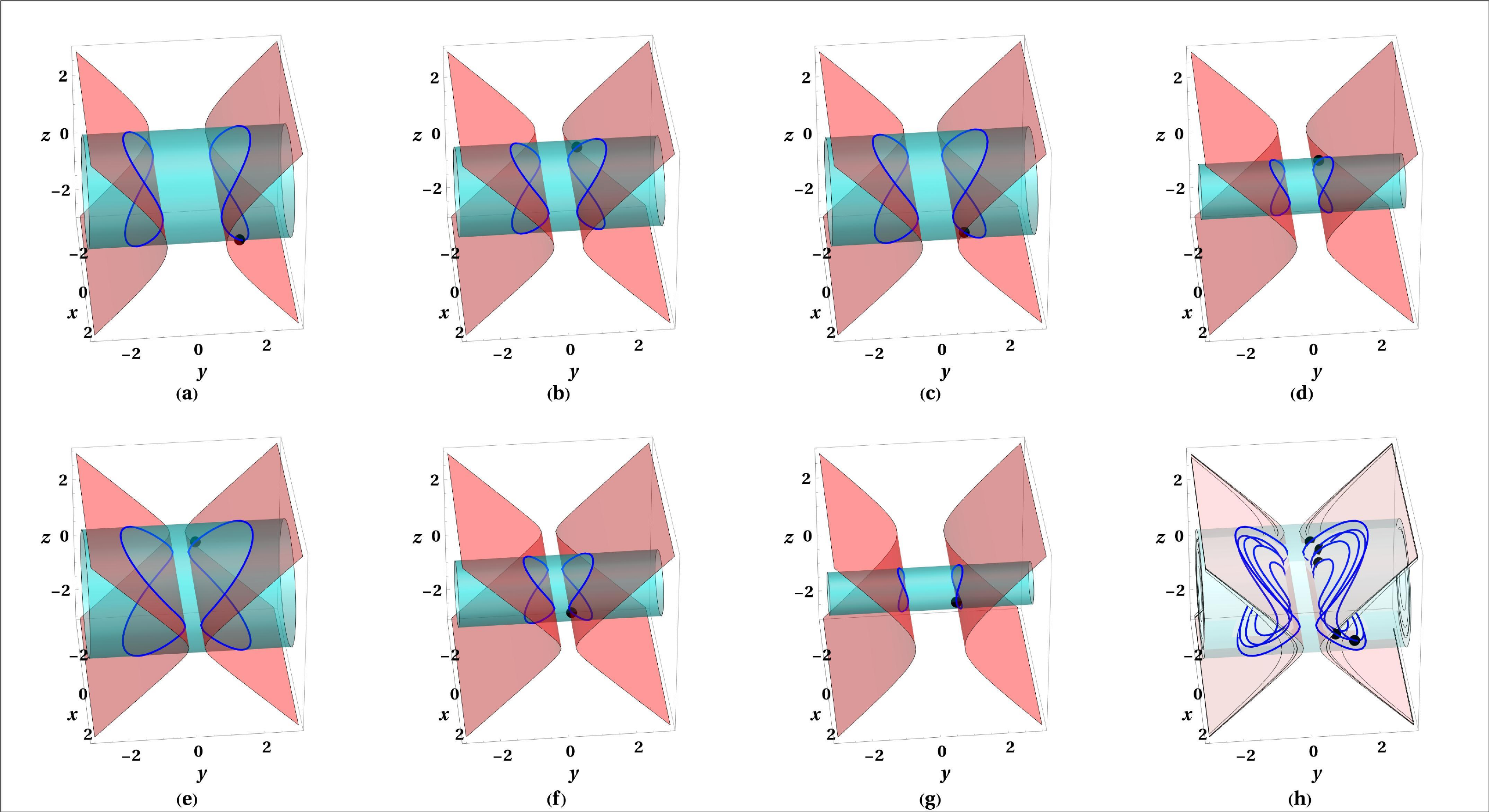}}
			\caption{Sub-figures (a) - (g)  depict the intersecting orbits of $H_1=H_1(t_0)$ and $H_2=H_2(t_0)$ surfaces with different values of $H_i(t_0)$ at time, $t_0=3, 20, 35, 48, 150, 300, 450$  of the four-wing chaotic attractor with parameters $(b,c,f)$=$(- 0.01, 1, -1 )$. The black colour dots represent the positions of the particle at time $t_0$. The sub-figure (h) shows the closely packed trajectories of four-wing attractor by considering the collection of intersecting orbits shown in (a) - (g). Different positions of black dots in (h) indicate hopping of the  phase point from one orbit to other  at different $t_0$, tracing the actual trajectory of chaotic attractor.}
			\label{collc2}
		\end{figure}

Equation (\ref{eq2}) is equivalent to the equation (\ref{NDeq}) and it implies that volume element of (u, v, w) phase space is conserved. Only difference between these two is that the former equation, in terms of new variables, defines non-dissipative trajectories with initial condition at different time instants and therefore, the equation  alone gives the full trajectory as the intersection of continuously changing Nambu surfaces. This equivalence establishes that Nambu mechanics can be extended to non-conservative Hamiltonians in the specific case when dissipation is linear. Under this condition, the doublet $H_{1}$ \& $H_{2}$ are no more constant of motion. They evolve according to the equation (\ref{DotS})(Notice that $H_1 = S$ for $\alpha = 1$ and $\beta =0$, while $S=H_2$ for $\alpha = 0$ and $\beta = 1$). The time evolution leads to a continuous change in relative orientation and area of the two surfaces. Therefore, the intersection of these surfaces, leading to a trajectory , does also change with time. The actual trajectory formed due to the intersection of continuously deforming surfaces can now be understood in the following manner: Because of the similarity of the equations (\ref{eq2}) and (\ref{NDeq}), we can still write equation of Nambu surfaces as $H_i = H_{i}(t_0)$, where $i=1,2$ and $t_0$ is an instantaneous initial time. Intersection of these equations gives a homoclinic orbit similar to the conservative case. For different $t_0$, we have distinctly different orbits. This is because of the change in size of the surfaces. The radius of the cylinder changes and the separation between the two sheets of the hyperboloid surface also changes with varying constant $H_i(t_0)$. This is demonstrated in FIG. \ref{collc2}(a)-(g). $H_{i}(t_0)$ are calculated from the time series i.e. using different ($x(t)$, $y(t)$, $z(t)$) at different time $t_0$. Several of these values are listed in TABLE \ref{table2}. For all these initial times, Nambu surfaces and their intersection are shown in FIG. \ref{collc2}. This figure shows that if we plot intersections for every instant $t_0$ in a single plot, we expect to get closely packed homoclinic orbits. Therefore, the actual trajectory of the full dynamical system is then the path traced by a phase point(represented by black colour dot) by hopping from one orbit to the adjacent orbit forming the desired attractor. This picture is depicted in FIG. \ref{collc2}(h) for a finite collections of homoclinic orbits. This mechanism also justifies the formation of a four wing structure in this case. Notice that one of the surface is a two sheeted hyperboloid. A single homoclinic orbit for a specific $t_0$ is formed by the intersection of one of the two sheets of the hyperboloid and the  cylindrical surface. This forms two wing orbit around two equilibria. When $H_i(t_0)$ varies with $t_0$, then the orbits corresponding to different values of $H_i(t_0)$ are distributed on both the sheets of the hyperboloid with different orientation and sizes. Therefore, while hopping from one orbit to other, the phase point traces four-wing trajectories as illustrated in  FIG. \ref{collc2}(h). Hence, we can clearly see how such a four-wing geometric attractor is formed purely because of specific geometry of surfaces, their relative orientations and radius of the cylindrical surface.
\begin{table}[htbp!]
	\centering
	\footnotesize
	\caption{\label{table2} Dynamical values of the Nambu doublet ($H_1$ \& $H_2$), calculated from the time series ($x(t)$, $y(t)$, $z(t)$) at several time instant $t_0$. }
	\addtolength{\tabcolsep}{1.8pt}
	\renewcommand{\arraystretch}{1.2}
	
	\begin{tabular}{ccc}
		
		\toprule[0.07cm]
		\midrule
		\multicolumn{3}{c}{\textbf{Four-wing system: ($a=0.2,b=-0.01,c=1,d=-0.4,e=-1,f=-1$), $IC=(1,1,1)$}}\\
		\midrule
		
		\bf{Time}  & \bf{($x,y,z$)} & \bf{$H_i=H_i(t_0), i=1,2$}\\    
		
		\midrule
		
		\multirow{1}{*} {$t=0 $} & $(1,1,1)$ 
		&$H_{1}=1,H_{2}=0.01$  
		\\ \addlinespace
		
		\multirow{1}{*} {$t=3 $} & $(1.134297,1.454767,-1.184064)$ 
		&$H_{1}=1.34420,H_{2}=-0.36901$  
		\\ \addlinespace
		
		
		\multirow{1}{*} {$t=20 $} & $(-1.220496,0.672179,0.521817)$ 
		&$H_{1}=0.880952,H_{2}=-0.0845482$  
		\\ \addlinespace
		
		\multirow{1}{*} {$t=35 $} & $(1.420197, 0.859965, -0.706224)$ 
		&$H_{1}=1.25786,H_{2}=-0.127457$  
		\\ \addlinespace
		
		\multirow{1}{*} {$t=48 $} & $(-0.720750, 0.580982, 0.390060)$ 
		&$H_{1}=0.335814,H_{2}=-0.088796$  
		\\ \addlinespace
		
		\multirow{1}{*} {$t=150 $} & $(-1.869253,0.442179, 0.368408)$ 
		&$H_{1}=1.81491,H_{2}=-0.0262149$  
			\\ \addlinespace
			
			\multirow{1}{*} {$t=300 $} & $(0.964893, 0.293572, -0.214692)$ 
			&$H_{1}=0.488556,H_{2}=-0.022193$  
			\\ \addlinespace
			
			\multirow{1}{*} {$t=450 $} & $(0.622904, 0.711482, -0.102801)$ 
			&$H_{1}=0.199289,H_{2}=-0.248848$  
			\\ \addlinespace
			
			\bottomrule[0.07cm]
		\end{tabular}
	\end{table}
	
We also repeated our calculation with different set of parameters. We observe similar geometry of Nambu surfaces, which yield four-wing attractor in these cases also. 

\subsection{\label{sec4d}Localization of four-wing attractor}
The concept of an attractor inherently requires it to be contained within a bounded region in the phase space. In other words, long time trajectory of a strange attractor fills a finite region and is localized within this region. We observe in the previous sections that the role of the conservative part of the velocity field is to produce a finite closed trajectory, which is the intersection of two conserved energy-like surfaces for a desired set of choice of the system parameters. We further observe that the linear dissipative part makes these surfaces dynamical, means they may expand or shrink due to dissipation. Now question arises what are the conditions that expansion and contraction remain within a finite region and their relative positions are such that their intersection is localized in the phase space. This question may be answered in two ways : (1) to show that there exists a surface which is localized and the attractor is enclosed within it, (2)  to show that intersection of the surfaces itself is confined within the two finite limits of intersection. The former was first introduced in \cite{doi:10.1080/02681119508806207} for Lorenz attractor, wherein several surfaces are proposed and shows that these surfaces are localized. Subsequently, the same idea is also applied by \cite{Roupas_2012}, where existence of such surfaces are derived using Nambu mechanics for Lorenz system. We apply here the first idea to show the localization of Nambu surfaces. We also apply the second idea to show that intersection of the two parent Nambu surfaces changes between two finite limits of intersections, showing the attractor localized. The second method is a new proposal.\\

The idea behind both the methods is the condition $\dot{S} = 0$ at a constant surface in the presence of dissipation. This means we need to find the extremum of the surface with respect to the variation in the state variables subject to the condition $\dot{S} = 0$. For this purpose, we apply Lagrangian multiplier method to find critical values of $x$, $y$, and $z$ such that $S=S_c(x_c,y_c,z_c)$ defines a constant surface.

\subsubsection{\label{sec4da}First Method: Identification of Localized Surfaces}

The various surfaces whose intersections give homoclinic orbits are already derived from Nambu mechanics and are given by the equation (\ref{S}). Since intersection lies at the surfaces, attractor is localized if a surface changes within a region of a constant surface. Therefore,  existence of a constant surface itself indicates localization. 

To apply Lagrange's multiplier method, we define a Lagrange function ($\mathcal{L}$) as
\begin{equation}\label{chaplang}
	\mathcal{L}(x,y,z,\lambda)=S(x,y,z)+\lambda\: \dot S(x,y,z)\Rightarrow \mathcal{L}=S+\lambda \:\dot S
\end{equation}
where $\lambda$ is Lagrange's multiplier.
Extremum values $x_c$, $y_c$ and $z_c$ are found by the condition
\begin{equation}\label{chapgrad}
	\vec\grad \mathcal{L}=0 \quad \Rightarrow \qty\Bigg(\pdv{\mathcal{L}}{x}, \;\pdv{\mathcal{L}}{y}, \;\pdv{\mathcal{L}}{z}, \;\pdv{\mathcal{L}}{\lambda})=0
\end{equation}

Using the expressions (\ref{S}) and (\ref{DotS}), we evaluate the expression (\ref{chaplang}) and find the following simultaneous equations from the  conditions (\ref{chapgrad}).

\begin{equation}\label{sol1}
	\begin{cases}
		x \;(1+2\lambda a)= 0 \\
		\eta f y\; (1+2\lambda d)= 0\\
		\left(\eta-\frac{c}{f}\right)\left( z-\rho\right)+2\lambda e \left(\eta-\frac{c}{f}\right) \left( z-\frac{\rho}{2}\right)	= 0\\ 
		ax ^{2}+\eta fdy^{2}+ e\left(\eta-\frac{c}{f}\right)\left(\left( z-\frac{\rho}{2}\right) ^2 - \left(\frac{\rho}{2}\right)^2\right)= 0 
	\end{cases}\,
\end{equation}

Solving equations (\ref{sol1}), we obtain four different values of $x_c$,$y_c$, $z_c$, and $\lambda $. These are listed in the TABLE \ref{T1}.
\begin{table}[!hbt]
	
	\centering
	\caption{\label{T1}A set of values of $S_c$ and corresponding $x_c$, $y_c$ and $z_c$ for the constant surfaces $S(x,y,z)=S_c$}
	\vspace{5pt}
	\scriptsize
	\begin{tabular}{|c|c|c|c|c| } 
		
		\toprule[0.04cm]
		\midrule
		
		\texttt{\textbf{x}} & \texttt{\textbf{y}} & \texttt{\textbf{z}}  & \texttt{\textbf{$\lambda$}} & \texttt{$\mathbf{S_c}$}\\
		\midrule
		$0$  & $ 0 $ & $ 0 $ & $-\frac{1}{e}$ & $0$ \\
		$0$ & $ 0 $ & $ \rho$ & $ 0 $ & $-\frac{1}{2}\left(\eta-\frac{c}{f}\right){\rho}^2$\\
		$0$ &  $\sqrt{\left(e-2d\right)\left(\eta-\frac{c}{f}\right)\frac{1}{\eta{fd}}\left(\frac{e\rho}{2\left(d-e\right)}\right)^2 }$ & $ \rho\frac{\left(2d-e\right)}{2\left(d-e\right)}$ &  $ -\frac{1}{2d}$ & $\frac{\left(\eta-\frac{c}{f}\right)}{8d\left(e-d\right)}\left(\rho{\left(e-2d\right)}\right)^2$\\
		$\sqrt{\left(e-2a\right)\left(\eta-\frac{c}{f}\right)\frac{1}{a}\left(\frac{e\rho}{2\left(a-e\right)}\right)^2 }$ &$0$& $ \rho\frac{\left(2a-e\right)}{2\left(a-e\right)}$ &  $ -\frac{1}{2a}$ & $\frac{\left(\eta-\frac{c}{f}\right)}{8a\left(e-a\right)}\left(\rho{\left(e-2a\right)}\right)^2$\\
		
		\bottomrule[0.04cm]
		
	\end{tabular}
\end{table}

The constant surface is then given by
\begin{equation}\label{SSc}
	\begin{aligned}
		S\left(x,y,z,\eta\right) = S_c \quad \Rightarrow \frac{1}{2}x ^{2}+\frac{1}{2}\eta fy^ {2}+ \frac{1}{2}\left(\eta-\frac{c}{f}\right)z^2 - {\eta}bz = S_c\\
		\frac{1}{2}x ^{2}+\frac{1}{2}\eta fy^ {2}+ \frac{1}{2}\left(\eta-\frac{c}{f}\right)\left(\left( z-\rho\right) ^2 - {\rho}^2\right) = S_c
	\end{aligned}
\end{equation}

It is noticed that by definition of a constant surface, $\dot{S}\left(x,y,z;\eta\right)$ at the surface  $S=S_c$ is zero. Whether the surface $S=S_c$ represents a localized region of the attractor or not is decided by the behaviour of $\dot{S}\left(x,y,z;\eta\right)$ near this surface. The expression for $\dot{S}\left(x,y,z;\eta\right)$ at and near the constant surface may be written by adding $e(S_c-S) =0$ to the equation (\ref{DotS}) so that $\dot{S}$ has quadratic terms only in $x$, $y$, and $z$ (eliminating the term ${\eta}ebz$). By doing so, we get the following expression for  $\dot{S}\left(x,y,z;\eta\right)$ at and near the constant surfaces:

\begin{equation}\label{DotSC}
	\dot{S}\left(x,y,z;\eta\right) =  \left(a-\frac{e}{2}\right)x ^{2}+\eta f\left(d-\frac{e}{2}\right)y^ {2}+ \frac{e}{2}\left(\eta-\frac{c}{f}\right)z^2 +eS_c	
\end{equation}

This equation satisfies the condition $\dot{S}\ge 0$ or $\dot{S}\le 0$ for all values of $x$, $y$, $z$ at or in the vicinity of the surface $S=S_c$ for the surface to be a localized region of the attractor . The first one is for repelling surface, where as the second condition for the attracting surface. Equal sign satisfy at the surface.

With the equation (\ref{DotSC}) for $\dot{S}$ near the constant surface, we can now analyse the localization property of the surface $S=S_c$ for the various types of Nambu surfaces discussed in the section \ref{sec4b}.

{\bf{Case-I: Cylindrical Surface :}} In this case $\eta = 0$ and $\frac{c}{f}<0$. We find $ S_c=0$ for all the four constant surfaces because $\rho = 0$ if $\eta=0$. Therefore, the equation (\ref{Scy}) with $S=S_c =0$ represents the localized cylindrical surface. This cylindrical surface and the equation (\ref{DotSC}) are respectively given by
\begin{equation}
	\frac{1}{2}x^{2}-\frac{1}{2}\frac{c}{f}z^{2}=0\quad,\quad   
	\dot{S}\left(x,y,z\right) =  \left(a-\frac{e}{2}\right)x ^{2}- \frac{e}{2}\left(\frac{c}{f}\right)z^2
\end{equation}
The first equation can only be satisfied for $x=0$ and $z=0$. Therefore, there exists no cylindrical localized surface  for this attractor.

{\bf{Case-II: Paraboloid Surface :}} In this case $\eta = \frac{c}{f}$. Here also $S_c =0$ for all the four constant surfaces. Replacing $S$ by $S_c$ in the equation (\ref{Spa}), we get the localized paraboloid surface. This surface  and the corresponding equation (\ref{DotSC})
are respectively given by
\begin{equation}
	\begin{aligned}
		\frac{1}{2}x ^{2}+\frac{1}{2} c y^ {2} -\frac{c}{f} b z = 0\\
		\dot{S}\left(x,y,z;\eta\right) =  \left(a-\frac{e}{2}\right)x ^{2}+c\left(d-\frac{e}{2}\right)y^ {2}
	\end{aligned}
\end{equation} 
Notice $\dot{S}\le 0$ is possible only when $d-\frac{e}{2}<0$ so that $y^2$ term is negative  while $x^2$ term is positive for $e<0$, $d<0$, $a>0$ and $c>0$ in our numerical result. Therefore, parabolic surface is a localized region under the above conditions.

{\bf{Case-III: Ellipsoid Surface :}} In this case $\eta > \frac{c}{f}$ and $\eta{f}>0$. Then $\eta$ has to be negative for ${f}<0$. For the present numerical study, $f$, $d$ and $e$ are negative and rest of the parameters are positive. Therefore $x^2$ term in the equation (\ref{DotSC}) remains positive, while $y^2$ term is negative for $d-\frac{e}{2}<0$ and $z^2$ term is also negative. The constant term $eS_c$ is negative for positive $S_c$ which is true for the third surface only and zero for the first surface. Therefore, $\dot{S}\le 0$ will be satisfied more likely for the third and the first surfaces.  The equation (\ref{Sel}) reduces to the following  localized surfaces. 

\begin{equation}\label{SelL}
	\frac{1}{2}x ^{2}+\frac{1}{2}\eta fy^ {2}+ \frac{1}{2}\left(\eta-\frac{c}{f}\right)\left(\left( z-\rho\right) ^2\right) = S_c+ \frac{1}{2}\left(\eta-\frac{c}{f}\right){\rho}^2
\end{equation}

{\bf{Case-IV: Hyperboloid Surface :}} While the above three surfaces are attracting, means an attractor is within the surface, but for hyperboloid, attractor remains outside the localized surface. In this case (i) $\eta =0$ with $\frac{c}{f} >0$, (ii) $\eta = \frac{c}{f}$ with $c<0$, (iii) $\eta{f}<0$, $\eta >\frac{c}{f}$, and (iv) $\eta <\frac{c}{f}$. For the present numerical result, first two are not applicable, while for the third, $\eta >0$ and for the last, $\eta <0$. In either case the localized surface is given by the equation (\ref{SelL}). The right hand side of this equation is positive for $\eta > \frac{c}{f}$, the first and third constant surfaces. Therefore, these surfaces are one sheeted hyperboloid, while for $\eta <\frac{c}{f}$, these two surfaces are two sheeted hyperboloid. In either case, the equation (\ref{DotSC}) satisfy the condition for localization.

Taking together, the ellipsoid and hyperbolic surfaces, an attractor is localised by the attracting compact surface and the repelling non-compact surface such that attractor remains inside a compact surface.

\subsubsection{\label{sec4db}Second Method: Localized Intersection of Surfaces}
A trajectory is the intersection of two surfaces defined by two Nambu Hamiltonians. One of the surface is a non-compact hyperboloid and the other is a compact one. Therefore, localization can be addressed by the simultaneous vanishing of the time derivative of these surfaces at the intersection, means for common $x$, $y$ and $z$. Since, in the presence of dissipation, surfaces are dynamical, the intersecting orbit is also dynamical and therefore, it traces a surface in the fractional dimensional space. This surface defines the boundary of the attractor and its geometry. We shall demonstrate this method considering only the parent surfaces $H_1$($\alpha =1$, $\beta =0$ in $S$) and $H_2$ ($\alpha =0$, $\beta = 1$ in S).

{\bf{Lorenz System}}: We first apply this new method to Lorenz system . The equation of the surface $S$ for the Lorenz attractor is taken from the reference \cite{Roupas_2012}. Surfaces $H_1$ and $H_2$ are found from $S$ for the Lorenz system as 

\begin{equation}\label{HH}
	H_{1}=\frac{Y^2}{2}+\frac{Z^2}{2}-\rho Z \quad \& \quad H_{2}=\rho Z - \frac{X^2}{2}
\end{equation}
And their corresponding time derivatives are
\begin{equation}
	\dot{H_{1}}= -Y^2-b\left(Y-\frac{\rho}{2}\right)^2+\frac{b\rho^2}{4}	\quad \&\quad \dot{H_{2}}=\sigma X^2-\rho b Z
\end{equation}

$X$, $Y$ and $Z$ are the respective scaled state variables \cite{Roupas_2012}, defined as

\begin{equation}
	x= X \quad ; \quad y= \sqrt{\frac{r}{\sigma}} Y\quad;\quad z= \sqrt{\frac{r}{\sigma}} Z \quad; \quad \rho=\sqrt{\sigma r}
\end{equation}

where, $x$, $y$ and $z$ are the original state variables and $\sigma$, $r$, and $b$ are the system parameters.

We now apply Lagrange's multiplier method to find constant surfaces $H_1 = H_{1c}$ and $H_2 = H_{2c}$ where, $H_{1c}$ and $H_{2c}$ are constants to be found from the conditions $\dot{H_{1}} = 0$ and $\dot{H_{2}} =0$. We first apply Lagrange's multiplier method to find extremum of $H_1$ and corresponding  $Y_c$ and $Z_c$. By putting these values in $\dot{H_{2}} = 0$, we find $X_c$. Therefore, at the intersection of $H_1 = H_{1c}$ and $H_2 = H_{2c}$, $\dot{H_{1}}$ and $\dot{H_{2}}$ simultaneously vanish. This means the trajectory itself is the localized boundary.

Hence , for $H_1$, Lagrange's function is
\begin{equation}
	\mathcal{L}(X,Y,Z, \lambda)=H_1+\lambda \:\dot{H_{1}}= \left(\frac{Y^2}{2}+\frac{Z^2}{2}-\rho Z \right)+\lambda \: \left(-Y^2-b\left(Y-\frac{\rho}{2}\right)^2+\frac{b\rho^2}{4}\right)
\end{equation}

The condition $\vec\grad \mathcal{L}=0$, gives possible values of $Y_c$, $Z_c$ and $\lambda$. Finally  $\dot{H_{2}} =0$ gives the value of $X_c$. Obtained values of these are listed in the TABLE \ref{T2}.
\begin{table}[!hpbt]
	\centering
	\caption{\label{T2}A set of values of $X_c$, $Y_c$, $Z_c$, $H_{1c}$ \& $H_{2c}$ for Lorenz attractor  such that simultaneously both $\dot{H_{1}}=0$ \&  $\dot{H_{2}}=0$.}
	\vspace{9pt}
	\begin{tabular}{|c|c|c|c|c|c| } 
		\toprule[0.04cm]
		\midrule
		\texttt{$\mathbf{X_c}$} & \texttt{$\mathbf{Y_c}$} & \texttt{$\mathbf{Z_c}$}  & \texttt{$\mathbf{\lambda}$} & \texttt{$\mathbf{H_{1c}}$} & \texttt{$\mathbf{H_{2c}}$}\\
		\midrule
		$0$  & $ 0 $ & $ 0 $  &  $\frac{1}{b}$ & $0$ & $0$ \\
		$\sqrt{\frac{b \rho^2}{\sigma}}$  & $ 0 $ & $ \rho $  & $ 0 $ & $-\frac{\rho^2 }{2}$ & $\frac{\rho^2 (2\sigma-b)}{2 \sigma}$ \\
		$\sqrt{\frac{\rho^2 b (2-b)}{2\sigma (1-b)}}$  & $ \sqrt{\frac{b^2 \rho^2 (b-2)}{4(1-b)^2}} $ & $ \frac{\rho (2-b)}{2(1-b)} $  & $\frac{1}{2}$ & $\frac{\rho^2 (b-2)^2}{8 (b-1)}$ & $\frac{\rho^2 (2-b)(2\sigma-b)}{4\sigma(1-b)}$ \\
		\bottomrule[0.04cm]
	\end{tabular}
\end{table}

One immediate conclusion may be drawn from the values of $X_{c}$ and $Y_{c}$ is that $b \geq 2$. The constant surfaces are now given by 
\begin{equation}\label{HHC}
	\frac{Y^2}{2}+\frac{Z^2}{2}-\rho Z = H_{1c}\quad \& \quad \rho Z - \frac{X^2}{2}=H_{2c}
\end{equation}
Since the equations (\ref{HHC}) are simultaneously satisfied at the intersection, we can find the equation of the intersection by replacing $\rho Z$ in the first by the second equation. Therefore, the equation of the intersection is
\begin{equation}\label{I}
	-X^2 + Y^2 + Z^2 = 2(H_{1c} + H_{2c} )
\end{equation}

From the TABLE \ref{T2}, we infer that $H_{1c} + H_{2c}$ has three values and therefore it is between two limits, $R_{min}$ and $R_{max}$. Therefore, the actual attractor is confined between the two boundaries whose geometry is given by the equation (\ref{I}). This method of localization gives precise geometry of the boundary, whereas the first method can only indicate the region of localization.\\
{\bf{Four wing System}}: The doublet are given by the equation (\ref{hh}) and the time derivative can be found from the equation (\ref{DotS}) as 
\begin{equation}
	\dot{H_1}= ax^2-\frac{c}{f}e z^2 \quad ,\quad \dot{H_2}=f d y^2+e z^2-bez
\end{equation}

Like in previous case, here also we define a Lagrange's function to find extremum of $H_1$ and $H_2$ subject to the conditions $\dot{H_1}=0$ and $\dot{H_2}=0$. We define Lagrange's function to find extremum for $H_2$.

\begin{equation}
	\mathcal{L}(x,y,z, \lambda)=H_2+\lambda \:\dot{H_{2}}= \left(\frac{1}{2}fy^2 +\frac{1}{2}z^2-bz\right)+\lambda \: \left(f d y^2+e z^2-bez \right)
\end{equation}
By equating $\vec\grad \mathcal{L}= 0$, we obtain
\begin{equation}\label{h21}
	\begin{cases}
		\pdv{\mathcal{L}}{x}= \text{Independent of x} \\
		\pdv{\mathcal{L}}{y}= fy(1+2 \lambda d)= 0\\
		\pdv{\mathcal{L}}{z}= (1+2 \lambda e)z - b (1+ \lambda e)= 0\\ 
		\pdv{\mathcal{L}}{\lambda}=\dot{H_2}=f d y^2 + e z^2-bez =0 
	\end{cases}\,
\end{equation}

By solving equation (\ref{h21}), we obtain a set of $(y_c, z_c, \lambda)$ values and the corresponding $H_{2c}$. These values also satisfy $H_1$ surface at the intersection. Hence, by substituting the obtained $z_c$ in the equation: $\dot{H_1}= ax^2-\frac{c}{f}e z^2 =0$; we find $x_c$ and corresponding $H_{1c}$.  These values are listed in TABLE \ref{T3}. It is clear from the TABLE \ref{T3} that the conditions $\frac{ce}{fa} \ge 0$, $e \ge 2d$ if $fd >0$ so that all values are real and valid.

\begin{table}[h]
	\centering
	\caption{\label{T3}A set of values of $x_c$, $y_c$, $z_c$, $H_{1c}$ \& $H_{2c}$ which satisfy $H_1 = H_{1c}$ \& $H_2 = H_{2c}$ simultaneously such that both $\dot{H_1}=0$ \&  $\dot{H_2}=0$.}
	\vspace{9pt}
	\begin{tabular}{|c|c|c|c|c|c| } 
		
		\toprule[0.04cm]
		\midrule
		
		\texttt{\textbf{x}} & \texttt{\textbf{y}} & \texttt{\textbf{z}}  & \texttt{\textbf{$\lambda$}} & \texttt{$\mathbf{H_{1c}}$} & \texttt{$\mathbf{H_{2c}}$}\\
		\midrule
		$0$  & $ 0 $ & $ 0 $  &  $-\frac{1}{e}$ & $0$ & $0$ \\
		$\sqrt{\frac{ceb^2}{fa}}$  & $ 0 $ & $ b $  & $ 0 $ & $\frac{cb^2 (e-a)}{2fa}$ & $-\frac{b^2}{2}$ \\
		$\sqrt{\frac{ce}{fa}\left(\frac{b(2d-e)}{2(d-e)}\right)^2}$  & $ \sqrt{\frac{(e-2d)}{fd} \left(\frac{be}{2(d-e)}\right)^2 } $ & $ \frac{b(2d-e)}{2(d-e)} $  & $-\frac{1}{2d}$ & $\frac{c (e-a) b^2 (e-2d)^2}{8fa (e-d)^2}$ & $\frac{b^2 (e-2d)^2}{8d(e-d)}$ \\
		
		\bottomrule[0.04cm]
		
	\end{tabular}
\end{table}
Replacing $z^2$ in the equation $H_2 = H_{2c}$ by $x^2$ using the equation $H_1 = H_{1c}$, we get the equation of the intersection representing the boundary of the attractor: 
\begin{equation}
	\frac{x^2}{c} + y^2 - \frac{2b}{f}z = \frac{2H_{2c}}{f} + \frac{2H_{1c}}{c}
\end{equation}
Since there are more than 1 values of $H_{1c}$ and $H_{2c}$, we get two limits of the right hand side of the above equation. Therefore, an attractor is localized between a lower and an upper limits of the phase space area.		

\section{\label{sec5}Conclusion}

The phase space plot from direct numerical solution to the four-wing three dimensional complex dynamical system with parameter values $a=0.2$, $b=-0.01$, $c= 1$, $d=-0.4 $, $e=-1$ and $f=-1$ agrees with the earlier studies. The attractor is found to have four lobes or wings like structure. Numerical diagnostic tools like bifurcation diagram, Lyapunov exponent and fractional dimension show the system is chaotic for the chosen parameter values. Stability analysis shows the system has five equilibrium points with one the central point, and the other four equilibria correspond to the origins of each wing of the four-wing attractor.

We successfully decompose analytically the velocity vector field into two parts, one the conservative  and the other linear dissipative. We find that decomposition is not unique but different decompositions do not affect the dynamical description within the framework of Nambu mechanics, the central part of the study. The same is also demonstrated for the Lorenz attractor. We are able to write the conservative part into two Nambu Hamiltonian like energy functions, which are the central functions for the geometric description of the attractor. The intersection of these two energy surfaces do resemble with the numerically found trajectory, demonstrating the validity of the application of Nambu mechanics to the study of chaotic system. We demonstrate that a similar treatment can be extended to the whole system, when dissipative part is included. The basic form of the system in terms of the two Hamiltonians remain the same when the dissipation is linear. But, in this case, the Hamiltonians are time dependent and intersecting orbits are dynamical, in the sense that their shapes and sizes change. We demonstrate that the collection of such dynamical orbits do form the actual attractor, demonstrating the validity and power of Nambu mechanics.

Since an attractor is a localized region in the phase space, we expect presence of several constant Nambu surfaces in the presence of dissipation, so that dynamical intersection of energy surfaces remain within a confined region. We demonstrate this property of the surfaces using two methods. The first method which is introduced earlier also in the case of Lorenz attractor, defines several possible surfaces from the combination of the two Nambu surfaces. This is possible because of canonical property of the Hamiltonians under canonical transformation. We identify four such surfaces, three attracting, namely cylindrical, paraboloid and ellipsoid and a repelling hyperboloid. Using the  Lagrange's method, we find that  these surfaces, except the cylindrical, provide localization of the attractor. The second method is more robust and provides an equation of the boundary of the attractor also. In this method, we find  a constant surface which passes through the intersection of the two Nambu surfaces. Since the intersection of the Nambu surfaces is the actual trajectory, therefore this surface equation defines the boundary of the attractor. While finding the equation of the intersection or constant Nambu surfaces, we are able to find conditions on the parameter values also from our analytical treatment. The following conditions are found which agree with our numerical study: $\frac{ce}{fa}\ge 0$, $e \ge 2d$ and $fd >0$. We also apply the second method to the Lorenz system and finds the condition $b>2$, agreeing with the earlier result.

Thus, Nambu mechanics theoretical framework, not only explain the geometric property of a multi-wing chaotic attractor , it can also find the localization and localized surface along with conditions on system parameters.

\appendix 
\section{\label{AA} Representation of Helmholtz-Hodge decomposition in terms of Nambu doublets}
\begin{itemize}
	\item We have, $\vec{A}= H_{1}\,\vec\grad{H_{2}} \quad\Rightarrow \vec\grad\cross{\vec{A}}= \vec\grad\cross{(H_{1}\,\vec\grad{H_{2}})}=(\vec\grad{H_{1}}\cross \vec\grad{H_{2}})+H_{1}\,(\vec\grad\cross{\vec\grad{H_{2}}})=\vec\grad{H_{1}}\cross \vec\grad{H_{2}}$ . The first term can not be written as $(\vec\grad\cross{H_{1}})\,\vec\grad{H_{2}}$
	because of violation of the rule as cross product can not be applied on a scalar $H_{1}$ . The second term $H_{1}\,(\vec\grad\cross{\vec\grad{H_{2}}})=0$, because curl of gradient of a scalar is always zero. Therefore, we can write
	\begin{equation}
	\vec{v}=\vec{v_{ND}}+\vec{v_{D}} = (\vec\grad\cross{\vec{A}})+ \vec{\grad} D= (\vec\grad{H_{1}}\cross \vec\grad{H_{2}})+\vec{\grad} D
	\end{equation}
	\item For $\vec{v_{ND}}=(\vec\grad{H_{1}}\cross \vec\grad{H_{2}})$\,, we can get $\vec\grad\cdot{\vec{v_{ND}}}=\vec\grad\cdot{(\vec\grad{H_{1}}\cross \vec\grad{H_{2}})}=0$ .\\
	As per the law of scalar triple product, the dot and cross product can be interchanged without changing the result. so, we can write $\vec a \dotproduct (\vec b \cross \vec c)=(\vec a \cross \vec b) \dotproduct \vec c \quad\Rightarrow \vec\grad\cdot{(\vec\grad{H_{1}}\cross \vec\grad{H_{2}})}=(\vec\grad\cross{\vec\grad{H_{1}}})\dotproduct \vec\grad{H_{2}}$ and $(\vec\grad\cross{\vec\grad{H_{1}}})=0$ as curl of gradient of a scalar is always zero.
\end{itemize}
\section{\label{BB}Derivation of Nambu doublets for two different decompositions of Lorenz system}
The Lorenz vector flow is given by 
	\begin{equation}
	\begin{cases}
		\dot{x}=\sigma (y-x) \\
		\dot{y}=x(r-z)-y\\
		\dot{z}=xy-bz
	\end{cases}\,
\end{equation}
\begin{itemize}
	\item Case-(i): Let us consider the following decomposition
	\begin{equation}
		\label{choice1}
		\vec v_{ND}=(\sigma y, x(r-z), xy) \quad, \quad \vec v_{D}=(-\sigma x, -y, -bz)\quad \quad\Rightarrow \vec v=\vec v_{ND}+\vec v_{D}
	\end{equation}
	Using ($\dot{x}, \dot{y}, \dot{z}$) of $\vec v_{ND}$ - part, let us construct Nambu doublet.
	$\frac{\dot{y}}{\dot{z}} =\frac{x(r-z)}{xy} \; \Rightarrow\: \frac{dy}{dz}= \frac{(r-z)}{y}\;\Rightarrow\:\int{y \;dy}= \int{(r-z)\; dz}\;\Rightarrow\: \frac{y^2}{2}=rz-\frac{z^2}{2}+const.\;\Rightarrow\:\frac{y^2}{2}-rz+\frac{z^2}{2}=const.\\
	\Rightarrow\:\frac{1}{2}[y^2-2rz+z^2]=const.\;\Rightarrow\:\frac{1}{2}[y^2-2rz+z^2+r^2-r^2]=const.\;\Rightarrow\:\frac{1}{2}[y^2+(z-r)^2]=const.\;(say\: H_{1})$
	Again, 	
	$\frac{\dot{z}}{\dot{x}} =\frac{xy}{\sigma y} \; \Rightarrow\: \frac{dz}{dx}= \frac{x}{\sigma}\;\Rightarrow\:\int{\sigma \;dz}= \int{x\; dx}\;\Rightarrow\:\sigma z=\frac{x^2}{2}+const.\;\Rightarrow\:\sigma z-\frac{x^2}{2}=const.\;(say\:H_{2})$
	\begin{equation}
		\label{decomp1}
		\begin{cases}
			H_{1}=\frac{1}{2}[y^2+(z-r)^2]=\text{Cylinder oriented along the x-axis} \\
			H_{2}=\sigma z - \frac{x^2}{2}=\text{Parabola oriented towards the +ve z-axis}
		\end{cases}\,
	\end{equation}
	$\vec\grad{H_{1}}=(\; 0, \;y,\; (z-r)\;)$  ,
	$\vec\grad{H_{2}}= (\; -x, \;0,\;\sigma \;)$ \quad and \quad $ (\vec\grad{H_{1}}\cross \vec\grad{H_{2}})= (\sigma y, x(r-z), xy)=\vec v_{ND}$
	\item Case-(ii): The second choice of decomposition is
	\begin{equation}
		\label{choice2}
		\vec v_{ND}=(-ry+\sigma y, -xz, xy) \quad, \quad \vec v_{D}=(-\sigma x+ry, -y+rx, -bz)\quad \quad \Rightarrow \vec v=\vec v_{ND}+\vec v_{D}
	\end{equation}
	Using ($\dot{x}, \dot{y}, \dot{z}$) of $\vec v_{ND}$ - part,\\	
	$\frac{\dot{z}}{\dot{x}} =\frac{xy}{-ry+\sigma y}\Rightarrow\ \frac{dz}{dx}= \frac{x}{(\sigma-r)}\Rightarrow\int{x \;dx}= \int{(\sigma-r)\; dz}\Rightarrow \frac{x^2}{2}=(\sigma-r)z+const.\Rightarrow\frac{x^2}{2}-(\sigma-r)z=const.\;(say\: H_{1})$\\
	Again, $\frac{\dot{y}}{\dot{z}} =\frac{-xz}{xy}\Rightarrow\ \frac{dy}{dz}= \frac{-z}{y}\Rightarrow\int{y \;dy}= \int{-z\; dz}\Rightarrow \frac{y^2}{2}=-\frac{z^2}{2}+const.\Rightarrow\frac{y^2}{2}+\frac{z^2}{2}=const.\;(say\: H_{2})$
	\begin{equation}
		\label{decomp2}
		\begin{cases}
			H_{1}=\frac{x^2}{2}-(\sigma -r)z=\text{Parabola oriented towards the -ve z-axis} \\
			H_{2}= \frac{1}{2}[y^2+z^2]=\text{Cylinder oriented along the x-axis} 
		\end{cases}\,
	\end{equation}
	$\vec\grad{H_{1}}=(\; x, \;0,\; -(\sigma-r)\;)$  ,
	$\vec\grad{H_{2}}= (\; 0, \;y,\; z\;)$ \quad and \quad $ (\vec\grad{H_{1}}\cross \vec\grad{H_{2}})= (-ry+\sigma y, -xz, xy)=\vec v_{ND}$
\end{itemize}
We noticed that in both set of choices of decomposition, one of the Nambu surface is always a cylinder and the other surfaces is always a parabola. Hence, we concluded that the Nambu doublet (a pair of Nambu Hamiltonians) of a system is always unique, irrespective of the decomposition choices.
\section{\label{CC}Derivation of Nambu Hamiltonians for the four-wing system }
From equation (\ref{nd}), we get  \quad $\vec v_{ND}=(cyz , bx - xz, fxy)$ and from this part, we can derive Nambu doublets.\\
$\frac{\dot{z}}{\dot{x}} =\frac{fxy}{cyz} \; \Rightarrow\: \frac{dz}{dx}= \frac{fx}{cz}\;\Rightarrow\:\int{cz \;dz}= \int{fx\; dx}\;\Rightarrow\:\frac{c z^2}{2} =\frac{fx^2}{2}+const.\;\Rightarrow\:\frac{fx^2}{2}-\frac{cz^2}{2}=const.\;(say\:H_{10})$\\
$\frac{\dot{y}}{\dot{z}} =\frac{bx - xz}{fxy}\:\Rightarrow \frac{dy}{dz}= \frac{(b-z)}{fy}\:\Rightarrow\int{fy \;dy}= \int{(b-z)\; dz}\:\Rightarrow \frac{fy^2}{2}=bz-\frac{z^2}{2}+const.\:\Rightarrow\frac{fy^2}{2}-bz+\frac{z^2}{2}=const. (say\: H_{20})$
But, we noticed that
\begin{equation}
	\vec\grad H_{10} \cross \vec\grad H_{20} = f \;v_{ND} \;\Rightarrow \frac{1}{f} * [\vec\grad H_{10} \cross \vec\grad H_{20}] = v_{ND} \;\Rightarrow (\frac{1}{f}*\vec\grad H_{10}) \cross \vec\grad H_{20}=v_{ND}\;\Rightarrow \vec\grad H_{1} \cross \vec\grad H_{2} =v_{ND} 
\end{equation}
Hence, we can write 
\begin{equation}
	\vec\grad H_{1}=\frac{1}{f} * \vec\grad H_{10}\;\Rightarrow H_{1}=\frac{1}{f} * H_{10}\;\Rightarrow H_{1}=\frac{1}{f} *\bigg[\frac{fx^2}{2}-\frac{cz^2}{2}\bigg]\qquad\Rightarrow H_{1}=\frac{1}{2}x^2-\frac{1}{2}\frac{c}{f}z^2
\end{equation}
\begin{equation}
	\vec\grad H_{2}=\vec\grad H_{20}\;\Rightarrow H_{2}= H_{20}\qquad\Rightarrow H_{2}=\frac{1}{2}fy^2+\frac{1}{2}z^2-bz
\end{equation}
Hence, the Nambu doublet for four-wing system (\ref{eq1}) for chosen parameter values ($b=-0.01, c=1, f=-1$) is found to be
\begin{equation}
	\label{doub1}
	\begin{cases}
		H_{1}=\frac{1}{2}x^2-\frac{1}{2}\frac{c}{f}z^2=\text{Cylinder oriented along the y-axis} \\
		H_{2}=\frac{1}{2}fy^2+\frac{1}{2}z^2-bz=\text{Hyperboloid having axis along  the x-axis }
	\end{cases}\,
\end{equation} 
\section{\label{DD} Construction of unique Nambu doublets from two different decompositions of four-wing system}
We can have more than one choice of decompositions for four-wing system also in such a way that $\vec v=\vec v_{ND}+\vec v_{D}$ and those are given by 
\begin{equation}
	\label{decompo1}
	\vec v_{ND}=(cyz, bx-xz, fxy) \quad, \quad \vec v_{D}=(ax, dy, ez)
\end{equation}
\begin{equation}
	\label{decompo2}
\text{and}\qquad \qquad	\vec v_{ND}=(cyz-by, -xz, fxy) \quad, \quad \vec v_{D}=(ax+by, bx+dy, ez)
\end{equation}
\begin{itemize}
\item Case-(i): For decomposition choice 1 given by equation (\ref{decompo1}), we have $\vec v_{ND}=(cyz, bx-xz, fxy)$ and the Nambu doublet is already derived in Appendix \ref{CC}.
\item Case-(ii): For decomposition choice 2 given by equation (\ref{decompo2}),\quad  $\vec v_{ND}=(cyz-by, -xz, fxy)$ and let us construct Nambu doublets out of this.\\
$\frac{\dot{x}}{\dot{z}}=\frac{cyz-by}{fxy} \Rightarrow\frac{dx}{dz}= \frac{(cz-b)}{fx}\Rightarrow\int{(cz-b) \;dz}= \int{fx\; dx}\Rightarrow\frac{c z^2}{2}-bz =\frac{fx^2}{2}+const.\Rightarrow \frac{1}{2}fx^2-\frac{1}{2}cz^2+bz=H_{10}$\\
	$\frac{\dot{y}}{\dot{z}} =\frac{- xz}{fxy}\:\Rightarrow \frac{dy}{dz}= \frac{(-z)}{fy}\:\Rightarrow\int{fy \;dy}= \int{-z\; dz}\:\Rightarrow \frac{1}{2}fy^2=-\frac{z^2}{2}+const.\:\Rightarrow\frac{1}{2}fy^2+\frac{z^2}{2}=H_{20}$
But, we find that 
\begin{equation}
	\vec\grad H_{10} \cross \vec\grad H_{20} = f \;v_{ND} \Rightarrow \frac{1}{f} * [\vec\grad H_{10} \cross \vec\grad H_{20}] = v_{ND} \Rightarrow (\frac{1}{f}*\vec\grad H_{10}) \cross \vec\grad H_{20}=v_{ND}\Rightarrow \vec\grad H_{1} \cross \vec\grad H_{2} =v_{ND} 
\end{equation}
Therefore, we get 
\begin{equation}
	\vec\grad H_{1}=\frac{1}{f} * \vec\grad H_{10}\;\Rightarrow H_{1}=\frac{1}{f} * H_{10}\;\Rightarrow H_{1}=\frac{1}{f} *\bigg[\frac{1}{2}fx^2-\frac{1}{2}cz^2+bz\bigg]\qquad\Rightarrow H_{1}=\frac{1}{2}x^2-\frac{1}{2}\frac{c}{f}z^2+bz
\end{equation}
\begin{equation}
	\vec\grad H_{2}=\vec\grad H_{20}\;\Rightarrow H_{2}= H_{20}\qquad\Rightarrow H_{2}=\frac{1}{2}fy^2+\frac{z^2}{2}
\end{equation}
Hence, for selected set of parameter values ($b=-0.01, c=1, f=-1$), the Nambu doublet is found to be
\begin{equation}
	\label{doub2}
	\begin{cases}
		H_{1}=\frac{1}{2}x^2-\frac{1}{2}\frac{c}{f}z^2+bz=\text{Cylinder oriented along the y-axis} \\
		H_{2}=\frac{1}{2}fy^2+\frac{z^2}{2}=\text{Hyperboloid having axis along  the x-axis }
	\end{cases}\,
\end{equation} 
\end{itemize}
From equations (\ref{doub1}) \& (\ref{doub2}), we observed that one of the Nambu surface is always a cylinder and the other surfaces is always a hyperboloid, though they have slightly different mathematical expressions for two different choices of decomposition. Therefore, we concluded that the Nambu doublet (a pair of Nambu Hamiltonians) of a system is always unique, even though the decomposition procedure is not unique.
\section{\label{EE}Derivation of dissipative Nambu mechanics part}
\begin{equation}\label{der15}
	x = e^{-\eta_1 t}u,\quad y = e^{-\eta_2 t}v,\quad  z = e^{-\eta_3 t}w 
\end{equation}
Now, 
\begin{equation}\label{der16}
	x = e^{-\eta_1 t}u \; \Rightarrow \: \dot{x}=\dot{u} e^{-\eta_1 t}-\eta_1 u e^{-\eta_1 t} \qquad \text{and}\qquad \pdv{}{x}=e^{\eta_1 t}\pdv{}{u}
\end{equation}
Similarly, we can write derivative for other variables $y$ and $z$.
Replacing old variables with new ones and using equation (\ref{der16}) in the above equation (\ref{DE}), we find 
\begin{equation}
	\label{der18}
	\begin{aligned}
		\dot{x}+{\eta}_1x= \dot{u} e^{-\eta_1 t}\\
		\dot{y}+{\eta}_2y= \dot{v} e^{-\eta_2 t}\\
		\dot{z}+{\eta}_3z= \dot{w} e^{-\eta_3 t}\\
	\end{aligned}
\end{equation}
Hence, equation (\ref{DE}) became 
\begin{equation}
	\label{der19}
	\begin{aligned}
		\dot{u} e^{-\eta_1 t}=\left(\vec\grad{H_{1}}\cross\vec\grad{H_{2}}\right)_{x} \\
		\dot{v} e^{-\eta_2 t}=\left(\vec\grad{H_{1}}\cross\vec\grad{H_{2}}\right)_{y} \\
		\dot{w} e^{-\eta_3 t}=\left(\vec\grad{H_{1}}\cross\vec\grad{H_{2}}\right)_{z} \\
	\end{aligned}
\end{equation}
We know 
\begin{equation}\label{der20}
	(\vec \grad{H_{1}}\cross\vec\grad{H_{2}})_{i}=(\vec \grad{H_{1}})_{j} \; (\vec \grad{H_{2}})_{k}-(\vec \grad{H_{1}})_{k} \; (\vec \grad{H_{2}})_{j} 
\end{equation}
So,
\begin{equation}
	\label{der21}
	\begin{split}
		(\vec \grad{H_{1}}\cross\vec\grad{H_{2}})_{x}&=\partial_{y}H_{1}\; \partial_{z}H_{2}-\partial_{z}H_{1}\; \partial_{y}H_{2}\\&
		=e^{\eta_2 t}\partial_{v}H_{1}\;\;e^{\eta_3 t} \partial_{w}H_{2}-e^{\eta_3 t}\partial_{w}H_{1}\;\;e^{\eta_2 t} \partial_{v}H_{2}\\&
		=e^{(\eta_2+\eta_3) t}\;\;\qty\big (\partial_{v}H_{1}\; \partial_{w}H_{2}-\partial_{w}H_{1}\; \partial_{v}H_{2})\\&
		=e^{(\eta_2+\eta_3) t}\;\;(\vec \grad_{q}{H_{1}}\cross\vec\grad_{q}{H_{2}})_{u}
	\end{split}
\end{equation}
With the help of equation (\ref{der21}), equation (\ref{der19}) can be rewritten as
\begin{equation}
	\label{der22}
	\begin{aligned}
		\dot{u}=e^{(\eta_1+\eta_2+\eta_3)t}\left(\vec\grad_{q}{H_{1}}\cross\vec\grad_{q}{H_{2}}\right)_{u} \\
		\dot{v}=e^{(\eta_1+\eta_2+\eta_3)t}\left(\vec\grad_{q}{H_{1}}\cross\vec\grad_{q}{H_{2}}\right)_{v} \\
		\dot{w}=e^{(\eta_1+\eta_2+\eta_3)t}\left(\vec\grad_{q}{H_{1}}\cross\vec\grad_{q}{H_{2}}\right)_{w} \\
	\end{aligned}
\end{equation}
transforming
\begin{equation}
	\label{der24} 
	d\tau=e^{-(\eta_1+\eta_2+\eta_3)t}{dt} \qquad
\end{equation}
we finally get
\begin{equation}
	\label{der26}
	\frac{d}{d\tau}(u,v,w)= \vec\grad_{\vec q} {H_{1}}\cross\vec\grad_{\vec q}{H_{2}}
\end{equation}
where,  $\vec q \equiv (u,v,w) = (e^{\eta_1 t}x,\; e^{\eta_2 t}y,\; e^{\eta_3 t}z)$ \quad and  \quad $H_{i} = H_{i}\left(x,y,z\right) = H_{i}\left(e^{-\eta_1 t}u, e^{-\eta_2 t}v, e^{-\eta_3t}w\right)$ , $i = 1$, $2$ .
Therefore, equation (\ref{der26}) confirmed that not only the conservative non-dissipative part, but also the whole dissipative chaotic system can be represented in terms of Nambu Hamiltonians. Hence, it is known as dissipative Nambu mechanics.
\bibliography{Reference}
\end{document}